\documentclass[letterpaper,11pt]{article}
\usepackage{jheppub_mod}
\usepackage{graphicx, amsmath, amssymb, amstext}
\usepackage{relsize, float, multirow}
\usepackage{array, epstopdf, hyperref, }
\usepackage{caption, subcaption, mathtools}
\usepackage{newtxmath}

\newcommand{\id}{{\rm d}}

\title{CMB anisotropy and BBN constraints on pre-recombination decay of dark matter to visible particles}
\author[a]{Sandeep Kumar Acharya,} 
\author[a]{Rishi Khatri}
\affiliation[a]{Department of Theoretical Physics, Tata Institute of 
Fundamental Research, Mumbai 400005, India}
\emailAdd{sandeepkumar@theory.tifr.res.in, khatri@theory.tifr.res.in}
\date{\today}
\abstract{ 
Injection of high energy electromagnetic particles around the recombination
epoch can modify the standard recombination history and therefore the CMB
anisotropy power spectrum. Previous studies have put strong constraints on
the amount of electromagnetic energy injection around the recombination era
(redshifts $z\lesssim 4500$). However, energy injected in the form of
energetic ($>$ keV) visible standard model particles is not deposited
instantaneously. The considerable delay between the time of energy
injection and the time when all energy is deposited to background baryonic
gas and CMB photons, together with the extraordinary precision with which
the CMB anisotropies have been measured, means that CMB anisotropies are
sensitive to energy that was injected much before the epoch of
recombination. We show that the CMB anisotropy power spectrum is sensitive
to energy injection even at  $z = 10000$, giving stronger constraints
compared to big bang nucleosynthesis and CMB spectral distortions. We
derive, using Planck CMB data, the constraints on long-lived unstable
particles decaying at redshifts $z\lesssim 10000$ (lifetime $\tau_X\gtrsim
10^{11}$s) by explicitly evolving the electromagnetic cascades in the
expanding Universe, thus extending previous constraints to lower particle
lifetimes. We also revisit the BBN constraints and show that the delayed
injection of energy is important for BBN constraints. We find that the
constraints can be  weaker by a factor of few to almost an order of
magnitude, depending on the energy, when we relax the
quasi-static or on-the-spot assumptions.
}
\notoc
\begin{document}
\maketitle
\newpage
\section{\label{sec:intro}Introduction}
  The incredible precision measurement of the cosmic microwave background (CMB) anisotropy power spectrum allows us to not just measure the 6 parameters of the standard $\Lambda$CDM cosmological model \cite{Pl2018} with high precision but also study extensions to it and constrain new physics beyond the standard model of particle physics. The CMB is sensitive to new physics such as dark matter self interactions \cite{SS2000,BKS2017}, dark matter-dark radiation interactions \cite{BMS2015,LMS2016}, decay of long-lived unstable particles \cite{ASS1998,CK2003}, evaporating primordial black holes \cite{H1975,SKLP2018}, and annihilation of dark matter to standard model particles \cite{Slatyer:2009yq,hutsi,Slatyer20161}. The physics in the dark sector can affect CMB both gravitationally \cite{DBK2014,ALM2014,LMS2016} and electromagnetically \cite{ASS1998,CK2003,Slatyer:2009yq,SKLP2018}. In particular, there are many particle physics motivated scenarios where dark matter can annihilate or decay to standard model particles \cite{W1982,G1983,MMY1993,FRT2003,FRT20031,CFM2002,ASS1998,AFSW2009,BW2012}, see \cite{F2010,BHS2005} for reviews.  
     \par
   \hspace{1cm}
   In this paper, we are interested in new physics that can inject
   electrons, positrons and photons with energy much greater than the CMB
   temperature into the primordial plasma. We will focus on the decay of
   dark matter ( or new long-lived unstable particles) but our general
   conclusions about the ability of CMB to constrain energy injection at
   high redshifts are applicable to other processes like evaporating black
   holes as well. Energetic electrons, positrons, and photons, injected by
   new physics, deposit their energy in the background baryon-photon plasma
   by heating background electrons and by ionization and excitation of
   neutral atoms. A fraction of energy escapes as low energy photons below
   10.2 eV (rest frame Lyman-alpha threshold)
   \cite{Slatyer:2009yq,Galli:2013dna} resulting in spectral distortion of
   the CMB. Increased ionization rate of neutral hydrogen and helium due to
   energy injection around recombination results in higher freeze-out
   number density of residual free electrons after recombination compared
   to the standard recombination \cite{Zks1969,Peebles1969}. Increased number of scatterings of CMB
   photons with free electrons damp the temperature anisotropy while giving
   a boost to polarization signal
   \cite{ASS1998,CK2003,Padmanabhan:2005es,Slatyer:2009yq,Galli:2009zc}. Precision measurement of the CMB anisotropy spectrum therefore puts constraints on the amount of electromagnetic energy that can be injected during the epoch of recombination and  hence on the parameters of the new physics such as the annihilation cross-section of dark matter, fraction of decaying dark matter and lifetime of decaying particles \cite{Slatyer20161,SC2017,PLS2017}, and abundance of primordial evaporating black holes as a function of black hole mass \cite{SKLP2018}. 
\par     
\hspace{1cm}
   Energy deposition for sub-keV electrons to a hydrogen and helium gas as
   a function of ionization fraction has been studied in
   \cite{SV1985,FS2010}. Monte Carlo simulation of higher energy electrons
   including   inverse Compton scattering
   (ICS) was done in
   \cite{VEF2010}. In these calculations, the high energy ($\gtrsim$ keV)
   electrons are evolved until their energy drops down to $\sim$eV, after
   depositing most of their energy to background particles, at which point
   the rest of their energy is deposited as heat. Keeping track of the
   evolution history from injected energy to $\sim$ eV energy for each
   injected particle makes these calculations computationally expensive. An
   alternative recursive solution to the above problem was provided in
   \cite{Slatyer:2009yq,KK2008,KKN2010}, who also included relativistic
   processes of electron-positron pair production and photon-photon elastic
   scattering. In this method, the pre-computed energy deposition history
   of lower energy particles is used to compute energy deposition of higher
   energy particles. Using this recursive method, the authors in
   \cite{SC2017} have calculated the constraints on fraction of decaying
   dark matter (with respect to total dark matter) for decay to
   monochromatic electron-positron and photon pairs as a function of dark
   matter mass ($m_X$) and lifetime ($\tau_X$) for  $\tau_X\gtrsim
   10^{13}~$s (or corresponding decay redshift  $z_X\lesssim 1200$) using
   Planck 2015 \cite{P20151} CMB anisotropy power spectrum data.  Similar
   constraints have been provided in \cite{PLS2017}, where the authors have
   used the publicly available results of energy deposition fractions from
   \cite{Slatyer20162} to constraint particle lifetimes  $\tau_X\gtrsim
   10^{12}~$s ($z_X\lesssim 4500$). The authors in \cite{PLS2017} also
   provide an effective on-the-spot ansatz, which means the energy injected at a
   particular redshift is immediately deposited at that redshift. They
   absorb the  beyond on-the-spot corrections into an effective energy
   injection history function. Their results
   show that the constraints on decaying dark matter abundance with this
   on-the-spot ansatz agrees well with the full calculations of
   \cite{SC2017} for  $\tau_X\gtrsim 10^{13}$s but not for lower lifetimes.   \par
   \hspace{1cm}
   The high energy particles that are injected into the plasma do not
   deposit all of their energy into heating, excitation, ionization and
   sub-10.2 eV photons instantaneously but over a period of time. This
   delay between the energy injection and deposition implies that the CMB
   anisotropies can be affected by energy that was injected much before the
   epoch of recombination. We have developed a new code for evolution of
   high energy particle cascades in the expanding Universe based on the
   method proposed in \cite{KK2008,KKN2010}. We calculate the constraints
   on dark matter decaying to monochromatic electron-positron and photon
   pairs by evolving the full electromagnetic cascade, starting from the
   initial keV-TeV high energy electrons, positrons and photons until eV
   energies, i.e. until  all of the energy has been deposited or escapes as
   CMB spectral distortions. Our code takes into account all relevant
   processes at all energies and thus presents a unified approach without
   separate treatments for the low and high energy parts of the cascade.
   We provide CMB anisotropy constraints, using Planck 2015 \cite{P20151}
   data, for lower lifetimes than what has been studied before, upto the
   point where the constraints from big bang nucleosynthesis (BBN) and spectral distortions become
   stronger. We will see below that the CMB anisotropies give the strongest
   constraints for energy injection at redshifts as high as
   $z\approx$10000. We will show this in the specific context of decay of
   long-lived unstable particles, but this conclusion will be true for any
   general energy injection scenario. 

The high energy photons in the cascade can also destroy primordial elements
produced in the BBN, changing their abundances
\cite{ENS1985,EGLNS1992,KM1995,KKMT2018,PS2015,HSW2018,FMW2019}. We revisit the BBN
constraints in section \ref{sec:bbn}. We will see that the delay between the energy injection and
deposition is important for the BBN constraints also and results in
weakening of constraints compared to the current constraints in literature
which use instantaneous deposition
approximation.

   \section{\label{sec:energydep}Physics of electromagnetic cascade in an expanding Universe}
Any injected electromagnetic energy is deposited in the fully ionized
baryonic gas as heat while for the partially neutral gas, a fraction of
energy goes into excitation and ionization of atoms. In addition, a
significant fraction of energy can escape as low energy photons with energy
less than 10.2 eV (Lyman-alpha threshold in the local rest frame)
\cite{Galli:2013dna}. These photons will show up as spectral distortion in the CMB  spectrum but take no part in the CMB anisotropy power spectrum analysis. We include the energy loss to these photons in the total deposited energy fraction to keep track of energy conservation.  
\par
\hspace{1cm}
Energy is deposited in the baryonic gas through various atomic collision
processes. A photon with energy, $E_{\gamma}>13.6$ eV, can photo-ionize a
neutral hydrogen atom creating a free electron with excess energy
$E_{\gamma}-13.6$ eV going into the kinetic energy of the free electron. This electron, if it has kinetic energy greater than 13.6 eV, can scatter with neutral atoms, ionizing them, and producing secondary free electrons or excite electrons in neutral atoms from ground to higher energy levels. It can also scatter with free electrons to deposit its kinetic energy as heat through Coulomb scattering. These atomic processes are efficient for an injected electron with energy $\lesssim$ keV. For a higher energy electron, ICS process becomes the dominant process. By this process the electron boosts CMB photons, producing a spectrum of high energy photons at the expense of its kinetic energy. These photons can then ionize neutral atoms or lose their energy by scattering with background electrons through Compton scattering. Compton scattering of high energy photons with bound electrons can transfer sufficient energy through electron recoil to ionize them. At even higher energy, photons can pair produce electron-positron pairs on scattering with background electrons, ions, neutral atoms and CMB photons or scatter elastically with CMB photons. These high energy particles create numerous energetic particles by boosting background particles. Thus, as the cascade progresses, the energy of injected particle is shared by more and more particles with energy of each particle in the cascade decreasing until all  of the energy is deposited as heat, excitation, ionization, or escapes as low energy photons. Thus, a high energy particle ($>>$ keV) deposits its energy by producing a lot of sub-keV secondary particles while direct deposition is negligible. 
 Since energetic electrons, positrons and photons are produced cyclically (
 photons boosting background electrons or pair-producing
 electrons-positrons and the high energy electrons and positrons boosting
 CMB photons), one has to evolve the electron, positron and photon spectra
 simultaneously. Electrons and positrons produced at one timestep can
 deposit their energy to baryonic gas or CMB photons at that timestep as
 their collisional rates are much higher compared to Hubble expansion rate
 ($H(z)$). But for photons, the collisional rates at high energy are
 comparable to Hubble rate for $z\lesssim 10000$
 \cite{Slatyer:2009yq,AK2018}. Therefore, we must evolve the photon
 spectrum with background expansion taken into account. More importantly,
 high energy photons deposit their energies on timescales much larger than
 the Hubble time ($t_H=1/H(z)$) and thus act as messengers of energy injection carrying information from redshifts as high as $z\approx$10000 to the recombination epoch, where they influence the CMB anisotropies. \par
\hspace{1cm}
A positron's energy loss mechanism is identical to that of an electron with the exception that it can annihilate with a background electron. Therefore, each injected positron can be approximated by an electron with the same kinetic energy plus two 511 keV photons \cite{Slatyer20162,AK2018}. 
\par 
\hspace{1cm}
From the above discussion, it is clear that the calculations can be
approximately  divided into low and high energy parts. For low energy, only
atomic processes\footnote{We club low energy Coulomb scattering also into
  atomic processes for this discussion} are important while for high
energy, atomic processes are negligible and only high energy relativistic
processes are important. The energy deposition of sub-keV particles to
background particles is a function of the ionization fraction of background
baryonic gas. This strategy, to  separate out the low and high energy
physics parts, was used in calculations of
\cite{Slatyer:2009yq,Slatyer20162}. They use tabulated results of energy
deposition fractions for sub-keV particles as a function of ionization
fraction of baryonic gas from monte-carlo calculations \citep{VEF2010},
which forms the low energy code. The high energy code evolves the particles
with all high energy scattering processes giving the spectra of low energy
particles at each timestep. These low energy particles are removed from
high energy code and fed to low energy code to calculate the energy
deposition fractions at that timestep. The interface between low energy to
high energy code was chosen to be 3 keV. The results are however sensitive
to this transition energy. Changing this interface slightly to $\sim$ 1 keV
makes a difference of few percent while changing it to 10 keV or few
eV can result in much bigger difference see (Fig. 1 and 2 of \cite{Slatyer20162}). \par  
\hspace{1cm}
In this work, we present a unified calculation to solve for energy
deposition fraction and evolution of electromagnetic cascade
simultaneously, taking into account all relevant physics without an
arbitrary division between low and high energy physics. In particular, we
do not rely on monte-carlo calculations in a static Universe of sub-keV
energy deposition, but evolve the sub-keV particles also in the expanding
Universe in the same code that evolves the high energy cascade.  We follow
the formalism of \cite{KK2008,KKN2010} and divide the energy range from eV
to TeV in 200 logarithmically spaced bins. The problem of electromagnetic
cascade in an expanding Universe then reduces to that of a system of
coupled ordinary differential equations for different particles and energy
bins. Energetic particles always lose energy to background particles and
move from high energy bins to lower energy bins. We, therefore, start by
first calculating the cascade of lowest energy bin and successively move to
higher energy bins. For injection of a particular particle, the subsequent
cascade of lower energy particles created by it, can be reused from previous
steps. The lower energy cascades computed once, are thus used over and over
again, thereby making these calculations fast. Interactions in-between high
energy injected particles can be neglected as their number density  is much
smaller compared to the number density of background particles. Formally,
we have reduced the problem of evolution of electromagnetic cascade to
solving a set of coupled linear algebraic equations which are triangular
and we are solving these equations by back-substitution
\cite{KK2008,KKN2010,AK2018}.  The cross-sections for various
high energy processes used in this paper are given in the appendix of
\cite{AK2018}. We give the cross-sections for atomic processes included in
this paper in  Appendix \ref{app:elec}. We refer the reader to \cite{AK2018} for
further technical details of the algorithm. We briefly review a few
important aspects of energy deposition in the next two subsections.
\par
\hspace{1cm}
  \par
 \hspace{1cm}          
      \begin{figure}[!tbp]
  \begin{subfigure}[b]{0.4\textwidth}
    \includegraphics[scale=0.8]{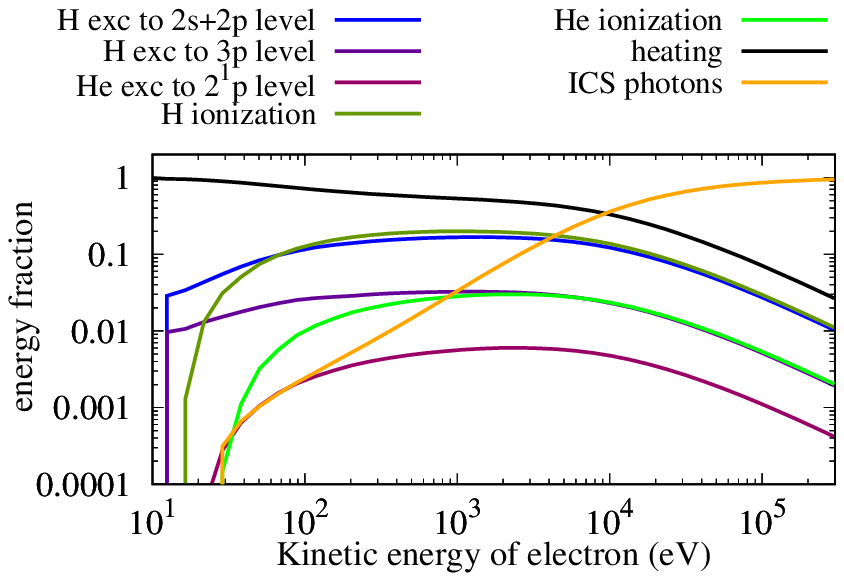}
    \caption{ $x_{\rm H}$=0.04, $x_{\rm He}\sim$0, z=1000.}
    \label{fig:depfracez=1000}
  \end{subfigure}\hspace{50 pt}
  \begin{subfigure}[b]{0.4\textwidth}
    \includegraphics[scale=0.8]{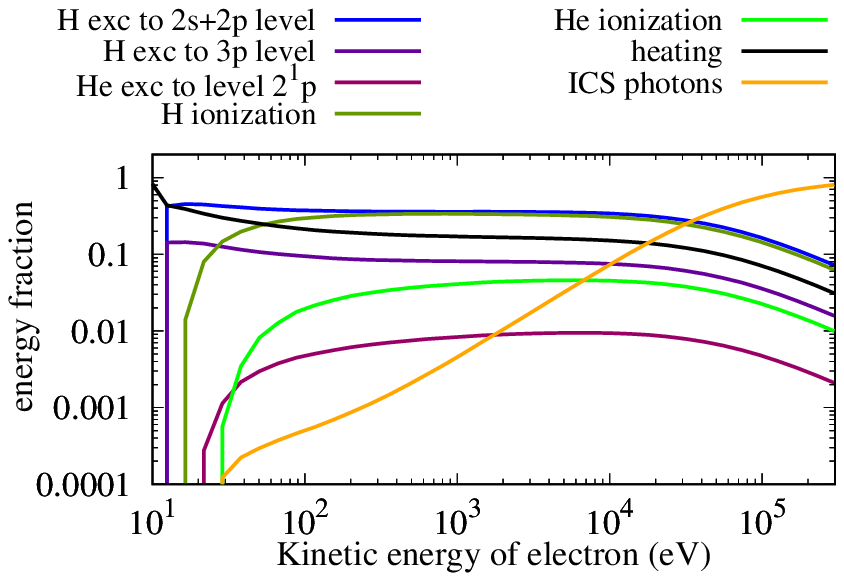}
    \caption{$x_{\rm H}$=0.0002, $x_{\rm He}\sim$0, z=100 }
    \label{fig:depfracez=100}
  \end{subfigure}
  \caption{Electron injection: Fraction of energy deposited through
    heating, excitation, ionization or CMB photons (Including sub-10.2 eV
    and higher energy photons) as a function of electron kinetic energy
    with the hydrogen ($x_{\rm H}$) and helium ($x_{\rm He}$) ionization fractions
    from {\bf Recfast++} \cite{Chluba2010}. Electron energy is deposited
    instantaneously as their energy loss rate is much faster than the
    Hubble rate (on-the-spot approximation). ICS photons include both
    sub-10.2 eV and higher energy photons. The photons with energy greater than 10.2 eV are processed in next timestep of the cascade evolution.}
  \label{eldepfrac}
\end{figure}
    \begin{figure}[!tbp]
  \begin{subfigure}[b]{0.4\textwidth}
    \includegraphics[scale=0.8]{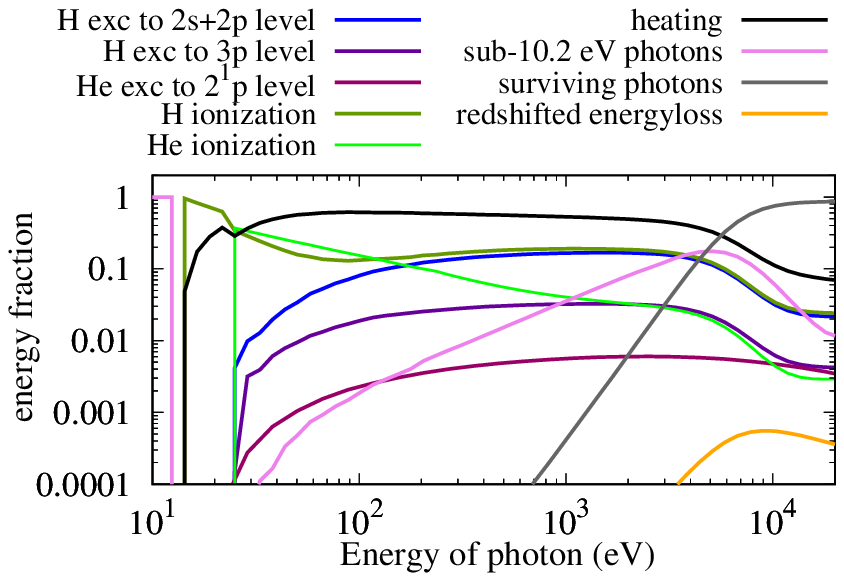}
    \caption{ $x_{\rm H}$=0.04, $x_{\rm He}\sim$0, z=1000.}
    \label{fig:depfracpz=1000}
  \end{subfigure}\hspace{50 pt}
  \begin{subfigure}[b]{0.4\textwidth}
    \includegraphics[scale=0.8]{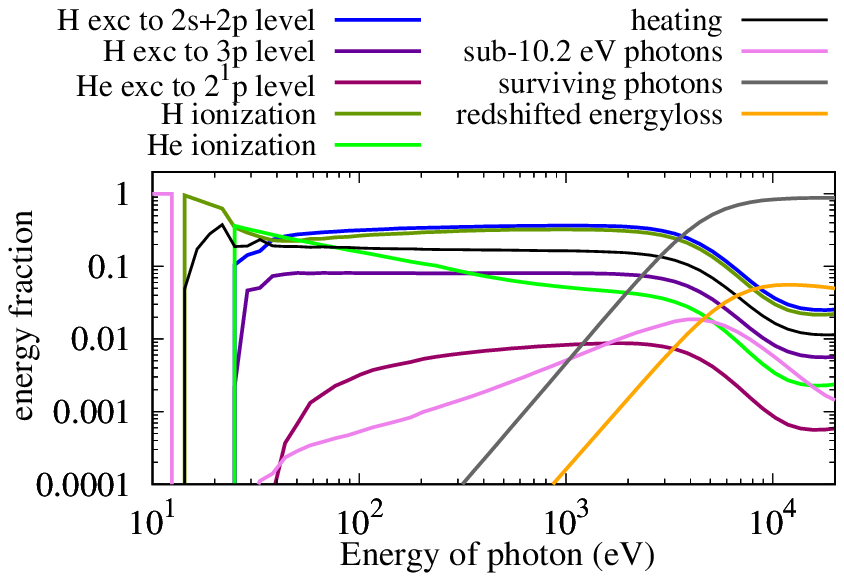}
    \caption{$x_{\rm H}$=0.0002, $x_{\rm He}\sim$0, z=100 }
    \label{fig:depfracpz=100}
  \end{subfigure}
  \caption{Photon injection: Fraction of energy deposited through heating,
    excitation, ionization and sub-10.2 eV photons, redshifted energy loss
    and surviving photons as a function of
    photon energy. }
  \label{pdepfrac}
\end{figure}
\subsection{\label{subsec:einj}Electron injection}
In Fig. \ref{eldepfrac}, we plot fractions of energy deposited instantly
(on the timescale much smaller than the Hubble time)  through heating,
excitation, ionization and to CMB photons by ICS as a function of the
kinetic energy of the injected electron at z=1000 and 100 (with the
ionization fractions calculated from \textbf{Recfast++}
\cite{Seager:1999bc,Chluba2010}). Helium ionization fraction is extremely
small at these redshifts. We see the deposited fractions as a function of
energy are somewhat flat in both plots at $\sim$ keV energies, explaining
the relative insensitivity of the choice of interface energy around
$\sim$keV between low and high energy code of \cite{Slatyer20162}. ICS is
not important  at low energies but becomes dominant at $\gtrsim$ 10
keV. The crossover between ICS and the atomic processes is redshift
dependent. We can see the crossover at z=100 ($\sim 10^4$ keV) to be at
higher electron kinetic energy than at z=1000 ($\sim 10^3$ keV). An
electron with Lorentz factor $\gamma$ can boost a photon with energy
$\epsilon$ to $\sim \gamma^2 \epsilon$ through ICS. Therefore, the energy
loss rate of an electron by ICS ($\propto \epsilon \propto 1+z$) is higher
at higher redshifts. For a partially ionized Universe, the dominant channel
of energy deposition is through heating. Fractions of energy deposited by
electrons through excitation to second level and ionization are similar
because the scattering cross-sections for these processes are of similar
magnitude. Excitation to higher energy levels becomes approximately one
order of magnitude smaller for each increase in level as the
cross-sections become smaller. Energy deposited to helium is smaller by a
factor of $\sim$ 10 compared to hydrogen just because the ratio of number
density of helium to hydrogen $f_{\rm He}\approx 0.08$, making helium an unlikely target for an incident electron. These qualitative statements are true even when the injected particles have energies of the order GeV-TeV, since these particles deposit their energy by first producing lots of $\sim$ keV secondary particles.
\par
\hspace{1cm}
\subsection{\label{subsec:pinj}Photon injection}
In Fig. \ref{pdepfrac}, we plot the fraction of energy deposited by a
photon to baryonic gas by atomic processes, energy-loss to sub-10.2 eV
photons, redshifted energy loss, and surviving photons  for energy
injection at $z$=100 and 1000 as a function of photon energy. These energy
fractions are calculated by first comparing the cooling rate of the injected photon,
including all scattering processes, with the Hubble rate at a
particular redshift, $H(z)$.
 The cooling time of photons  is given by, $t_{\rm cool}=1/(d\ln E_{\gamma}/dt)$
 \cite{Slatyer:2009yq,AK2018}, where $\id\ln E_{\gamma}/\id t$ is the total
 energy loss rate of photons due to scattering with the background
 particles {and $t$ is the proper time.} Therefore, the fraction of energy lost and deposited by a photon,
 ionizing neutral gas and boosting free electrons, is given by the ratio
 $\frac{t^{-1}_{\rm cool}}{t^{-1}_{\rm cool}+H(z)}$, while rest of the energy
 survives. The free electrons created  during these processes deposit their
 energy immediately according to the discussion in the previous
 section. Since a large fraction of the photon energy is deposited by first boosting
 electrons, it explains the similarity of the energy deposition fractions by
 electrons and photons for $\sim$keV energies. Close to the threshold
 of ionization of neutral hydrogen (13.6 eV) and neutral helium (24.6 eV),
 the dominant fraction of energy goes into ionization. At higher
 energy, a photon has a higher chance to survive since the high energy
 collision rates for photons and the Hubble rate are of the same
 order. These surviving photons then redshift to the next timestep.
   \section{Modification to the recombination history from electromagnetic energy injection}
   The problem of cosmological recombination has been studied in great
   detail beginning with the first calculations of
   \cite{Zks1969,Peebles1969}. The need for multilevel calculations was
   emphasized by \cite{seager2000} for precision calculations needed by
    the CMB experiments in the 21st century. There has been tremendous
    progress since then, culminating in the fast effective multilevel recombination
    codes
    \cite{CS2006,SH2008,RCS2008,Chluba2010,GH2010,CVD2010,AH2010,Hh2011}. Most of the complicated dynamics of recombination can be captured by the effective 3-level atoms originally proposed by \cite{Zks1969,Peebles1969}, with the equations suitably modified by a fudge factor \cite{Seager:1999bc} or a fudge function \cite{Chluba2010}. In these calculations, only the first excited levels with the principal quantum number $n=2$  of hydrogen and helium are resolved, in addition to the ground state and the continuum or the ionized state of the atom. However, the energetic electrons in the electromagnetic cascade can also excite the hydrogen and helium atoms to $n=2$ as well as higher energy levels. To take these excitations into account requires us to resolve the higher levels. The energy injection modifies the standard recombination calculations by adding  extra source or sink terms to the differential equations governing the population of different levels \cite{C2010}. In particular, the equations for the ground states of hydrogen and helium get an extra sink term due to direct ionizations from the ground state,
\begin{align}
\left.\frac{\id x_{\rm H_{\rm 1s}}}{\id t}\right|_{\rm inj}&=
-f_{\rm Hi}(z)\frac{\id E_{\mathrm{inj}}/\id t}{n_{\rm H} E_{\rm H_{\rm 1s}}}\\
\left.\frac{\id x_{\rm He_{\rm 1s}}}{\id t}\right|_{\rm inj}&=
-f_{\rm Hei}(z)\frac{\id E_{\mathrm{inj}}/\id t}{n_{\rm H} E_{\rm He_{\rm 1s}}},
\end{align}
where $x_{\rm H_{\rm 1s}}=n_{\rm H_{\rm 1s}}/n_{\rm H}$, $x_{\rm He_{\rm
    1s}}=n_{\rm He_{\rm 1s}}/n_{\rm H}$, $n_{\rm H}$ is the total (ionized
+ neutral) hydrogen number density, $n_{\rm H_{\rm 1s}}$ and $n_{\rm
  He_{\rm 1s}}$ are the number densities of neutral hydrogen and helium
atoms respectively,  $E_{\rm H_{\rm 1s}}$ and $E_{\rm He_{\rm 1s}}$ are the
binding energies for hydrogen and helium respectively,
$dE_{\mathrm{inj}}/dt$ is the rate of injection of energy
density,  and $f_{\rm Hi}$ and $f_{\rm Hei}$ are the fractions of injected
energy going into hydrogen and helium ionizations respectively. Similar
terms are also added for excitations to the equations for the levels
involved in the respective transitions
\cite{C2010}. Note that we do not assume that the ionization energy is
divided among the hydrogen and helium according to their relative number
density but calculate the respective fractions explicitly during the
cascade evolution.
 The energy deposition fractions $f_i(z)$, where $i$ labels the different
 channels such as hydrogen and helium ionization, excitations etc,  depends
 upon the redshift of injection, the energy of the injected particle, and
 whether the injected particle is an electron, positron or photon. \par
  \hspace{1cm}
  We also take into account the heating of background baryonic gas due to
  energy injection by adding a source term to baryon temperature ($T_{\rm
    b}$) evolution equation \cite{Zks1969,Peebles1969,Galli:2013dna},
  \begin{equation}
  (1+z)\frac{\id T_{\rm b}}{\id z}=\frac{8\sigma_{\rm T} a_{\rm R}
    T^4_{\rm CMB}}{3m_{\rm e} cH}\frac{x_{\rm e}}{1+f_{\rm He}+x_{\rm
      e}}(T_{\rm b}-T_{\rm CMB})-\frac{2}{3k_{\rm B}H}\frac{K_h}{1+f_{\rm He}+x_{\rm
      e}}+2T_{\rm b},
  \end{equation}
  where $x_{\rm e}=n_{\rm e}/n_{\rm H}$, $n_{\rm e}$ is the free electron
  number density, $\sigma_{\rm T}$ is the Thomson cross section, $T_{\rm
    CMB}$ is the CMB temperature, $a_{\rm R}$ is the radiation constant,
  $m_{\rm e}$ is the mass of electron, $k_{\rm B}$ is the Boltzmann
  constant, $c$ is the speed of light, 
  $K_h=f_h\frac{(\id E_{\mathrm{inj}}/\id t)}{n_{\rm H}}$ parameterizes the
  heating from extra energy injection, and $f_h(z)$ is the fraction of
  injected energy used up in heating the baryonic gas. At $z\gtrsim  200$, the gas
  temperature and the CMB temperature are efficiently coupled due to
  Compton scattering between  residual free electrons and CMB photons. As the CMB photons
  outnumber baryons by more than nine orders of magnitude, increasing the
  CMB temperature by even a small amount requires huge amount of energy
  injection. Since  the fraction of energy going into ionization of neutral
  hydrogen and heating of baryonic gas for a high energy particle injection
  are of the same order, {as shown in Figs. \ref{eldepfrac} and \ref{pdepfrac}}, this scenario would ionize the whole Universe, and is  thereby ruled out by current data. We, therefore, expect the correction to baryonic temperature evolution to be of negligible importance  compared to the ionization of neutral gas for CMB anisotropies. However, the corrections to baryonic temperature evolution can be important for 21cm cosmology and reionization \cite{LRS2019}. \par
  \hspace{1cm}
  Electromagnetic energy injection can ionize singly ionized helium as
  well. The recombination of doubly ionized helium to singly ionized
  helium is given by Saha equilibrium solution to a very good approximation
  and this approximation is implemented in recombination codes to make the
  codes faster. {Since the first recombination of helium happens
    very early ($z\approx 6000$), when the hydrogen is fully ionized,
    the extremely large Thomson optical depth makes the visibility function
    \cite{Zks1969,Peebles1969} vanishingly small. We can therefore ignore
    the extra ionizations of helium when calculating the CMB anisotropies. However, the
  extra ionizations and deviation from the Saha solution are} important for
  recombination line spectral distortions \cite{CA2016}. We should 
  clarify that we take both  neutral and singly ionized helium into account
  when evolving the electromagnetic cascades {and calculating the energy
   deposit fractions} as explained in the previous section.    
   \begin{figure}
\centering
\includegraphics[scale=1.0]{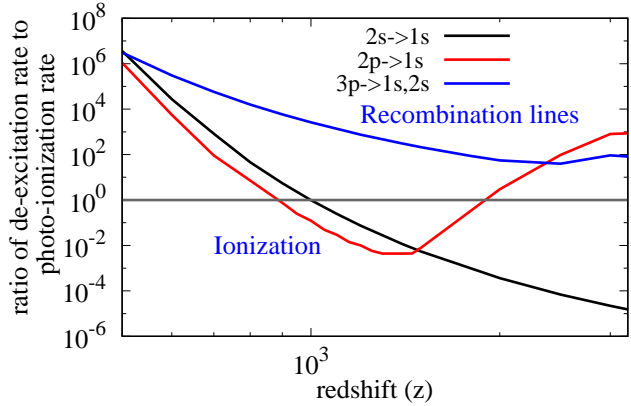}
\caption{Ratio of \emph{net} de-excitation rate (2s, 2p $\rightarrow$ 1s, 3p
  $\rightarrow$ 1s, 2s, {after taking into account the  escape probability}) to photo-ionization rate from excited levels of
  hydrogen as a function of redshift. We have labeled regions where
  the contribution to ionization is important vs where most of the excitation
  energy goes into recombination lines.}
\label{fig:ratio}
\end{figure}

\subsection{\label{subsec:exc}Effects of excitations}   
\begin{figure}
\centering
\includegraphics[scale=1.0]{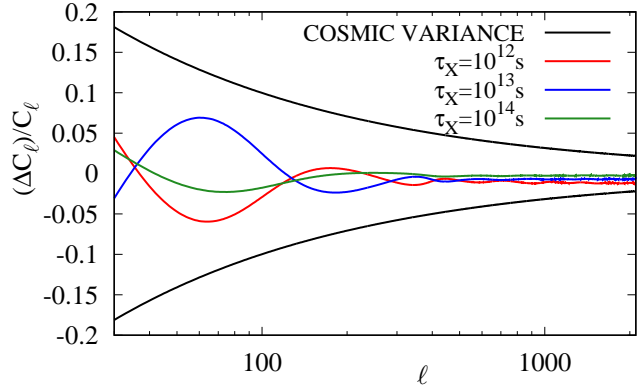}
\caption{{Fractional difference in the CMB angular power spectrum $\Delta
  C_{\ell}/C_{\ell}$, where $\Delta C_{\ell}$ is the difference in angular
  power spectrum $C_{\ell}$ calculated using  \textbf{Recfast++} (only
  direct ionizations from the energy injection included) and
  \textbf{CosmoRec} (including also extra excitations)
  for 1 TeV $e^-e^+$ injection. We show three different lifetimes and choose $f_X$ to be
  the upper limit (derived in this work) for the corresponding lifetime. Also shown is the cosmic
  variance $\sqrt{2/(2\ell+1)}$. }}
\label{fig:cosmovsrec}
\end{figure}
    The extra ionizations from energy injection can have two
    contributions: (i) direct ionization from the ground
    level and (ii) excitation to higher levels and subsequent {photo-ionization} by
    CMB photons. As we saw in the previous section, the energy fraction
    going into excitation is comparable to the energy fraction going into
    direct ionization from the ground state. An excited atom will however
    not necessarily get ionized. It can also de-excite to a lower level,
    emitting a recombination line photon which can escape or, if it is a
    Lyman-series photon, be reabsorbed.  We plot, in Fig. \ref{fig:ratio}, the
    ratio of {\emph{net} de-excitation rate} $P_i^{\mathrm{esc}}A_i$, where
    $P_i^{\mathrm{esc}}$ is the escape probability for the emitted photon
    \cite{Zks1969,Peebles1969,Seager:1999bc,Chluba2010,Hh2011} and $A_i$ is
    the spontaneous transition rate, to the ionization rate for different
    levels $i$ as a function of redshift. For the first excited levels of
    hydrogen, photo-ionization dominates de-excitation at $z\gtrsim
    1000$. At these redshifts, however, free electrons are likely to
    recombine with protons due to their high number density.  There is, thus,
    almost no modification to the recombination history making the CMB anisotropy
    power spectrum insensitive to extra ionizations from the energy that is absorbed at these high
    redshifts. At lower redshifts, the CMB power spectrum is sensitive to the
    increase in the residual electron fraction due to extra ionizations. However at $z\lesssim 1000$, photo-ionization from the
    first few levels becomes much less likely compared to de-excitation as
    the  CMB temperature decreases, as seen in  Fig. \ref{fig:ratio}. We, therefore, do not expect excitations to be important in
    increasing the post-recombination ionization fraction. Hence,
    extra excitations from energy injection are not important for CMB
    anisotropies. Most of the energy going into excitations  will
    therefore go
    into the cosmological recombination spectrum.

We check this explicitly using
    \textbf{CosmoRec} code \cite{Chluba2010} to take into account
    contribution to excitations upto third level for hydrogen and second
    level of helium by explicitly solving for the excited states without the
    assumption of effective three level atom.  We have used \textbf{CosmoRec} setting with 10 hydrogen and
    helium interface shells \cite{AH2010}, 500 hydrogen levels  and all of radiative transfer effects flags turned
    on. {We include contribution from injected energy to excitations from
    the ground state to 2s, 2p, and 3p
    levels of hydrogen and $2{}^1{\rm p}$ level of helium. The fractional difference
    in the CMB angular power spectrum $C_{\ell}$ as a function of multipole
    $\ell$ between \textbf{CosmoRec} (with excitations) and
    \textbf{Recfast++} (without  excitations) is shown in
    Fig. \ref{fig:cosmovsrec} and compared with the cosmic variance. We
    show the comparison for pre-recombination decay ($\tau_X=10^{12}~{\rm
      s}$), decay during recombination ($\tau_X=10^{13}~{\rm s}$), and a post
    recombination case ($\tau_X=10^{14}~{\rm s}$) using  the 
     corresponding upper limits for decay fraction $f_X$ derived below
    in this work. We find that the difference is much smaller than the
    cosmic variance.}
  
    Henceforth, we use \textbf{Recfast++}, since it is much faster, with contribution from the energy
    injections to only direct ionizations included {while calculating the
    recombination history. }
    \begin{figure}[!tbp]
  \begin{subfigure}[b]{0.4\textwidth}
    \includegraphics[scale=1.0]{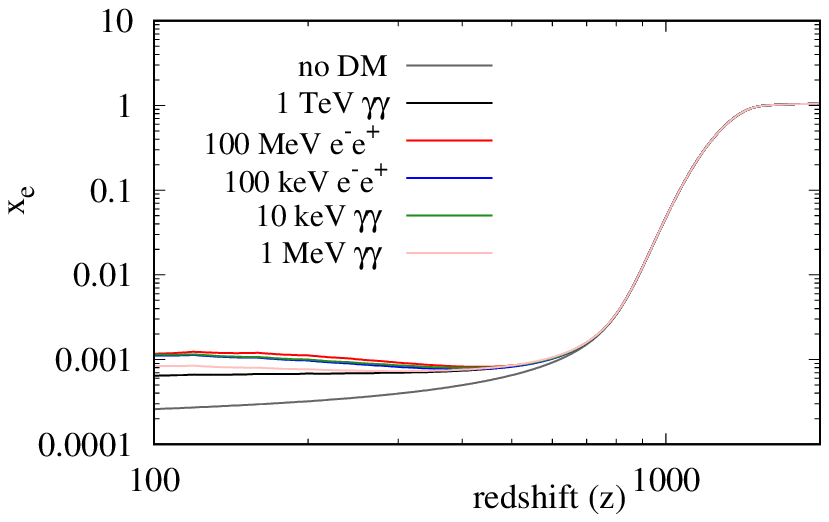}
     \caption{ }
     \label{fig:xetau141}
  \end{subfigure}\hspace{65 pt}
  \begin{subfigure}[b]{0.4\textwidth}
    \includegraphics[scale=1.0]{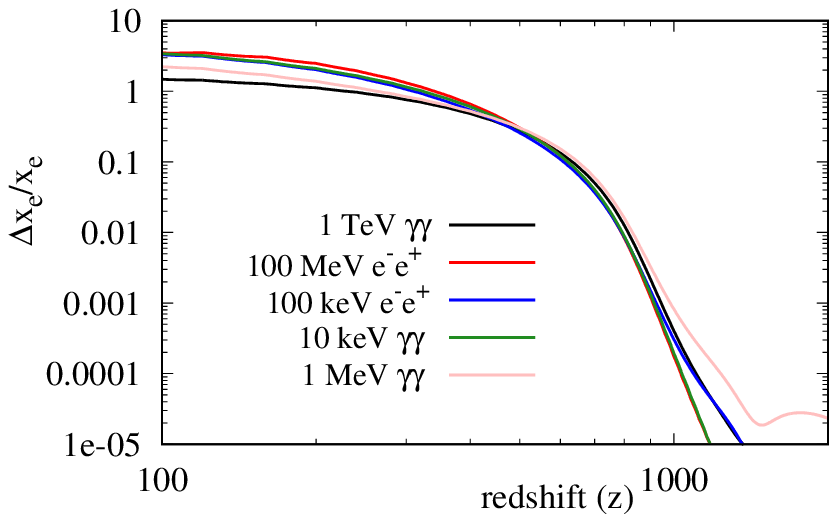}
     \caption{}
    \label{fig:xetau14}
    \end{subfigure}\\
    
    \begin{subfigure}[b]{0.4\textwidth}
    \includegraphics[scale=1.0]{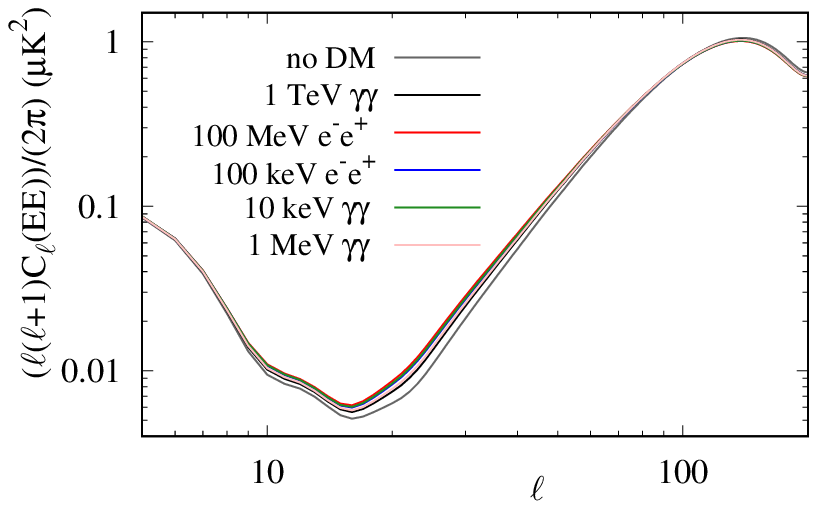}
     \caption{ }
  \label{fig:ClTTtau14}  
  \end{subfigure}\hspace{65 pt}
  \begin{subfigure}[b]{0.4\textwidth}
    \includegraphics[scale=1.0]{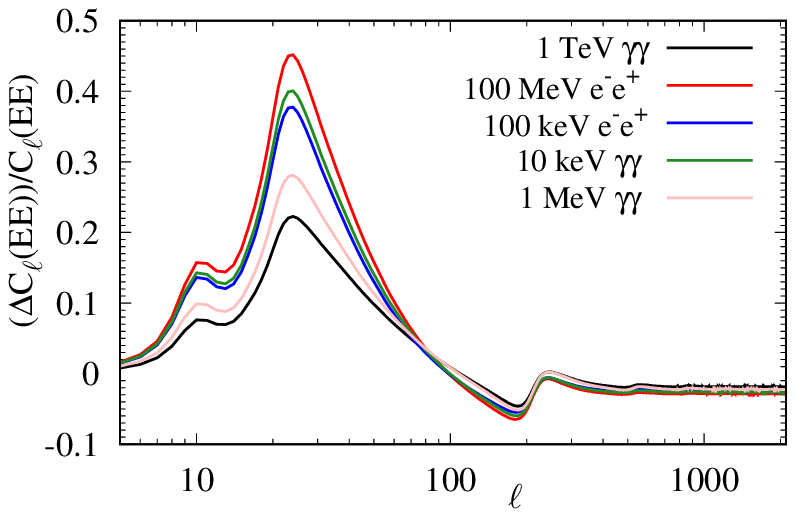}
     \caption{}
   \label{fig:ClEEtau14}   
   \end{subfigure}
   \caption{(a) Recombination history for different energy injection scenarios, (b) fractional change in ionization fraction, (c) CMB E-mode polarization spectrum, (d) fractional change in CMB E-mode power spectrum. All plots are for dark matter decay with lifetime ($\tau_X$)=$10^{14}$s and fraction of decaying dark matter ($f_X$) is equal to the 2-$\sigma$ upper limits derived in this paper.}
  \label{fig:tau14}
  \end{figure}
\par    
\hspace{1cm}
To illustrate the main effects on the CMB anisotropies, we plot in Fig. \ref{fig:tau14} the recombination histories and the E-mode
polarization power spectrum for energy injection from dark matter decay
with lifetime ($\tau_X=10^{14}{\rm s}$) longer than recombination
era. The main effect is a boost in the residual electron fraction, which
changes by order unity or more, after recombination. We expect this change
to affect modes entering the horizon between the recombination and the
reionization epoch. This is most clearly and dramatically seen in the E-mode
polarization spectrum. The E-mode polarization spectrum gets a boost for
the mulitpoles ($\ell$)  between the first recombination peak and the
reionization bump, with fractional change as large as $50\%$. However, the
absolute signal for these multipoles is still very small, because the
ionization fraction $x_{\rm e}\ll 1$. Therefore, the CMB constraints will be
driven by not these obvious signals but by more subtle changes in the rest
of the CMB temperature and polarization power spectrum, where the actual
signal is much higher. In particular, the percent level suppression, due to
extra optical depth given by the enhanced residual electrons, of the
high $\ell$ modes is important.  We give the plots for smaller lifetimes as well as the
CMB temperature power spectrum in Appendix \ref{app:plots}.
\color{black}
\section{\label{sec:energydepfull}Energy injection and deposition from dark matter decay}
Energy injection from dark matter decay can be parameterized by,
\begin{equation}
\frac{\id E_{\mathrm{inj}}}{\id t}=\frac{f_X}{\tau_X} \rho_c c^2 (1+z)^3 e^{-t/\tau_X},
\label{dmeq}
\end{equation}   
    where $f_X$ is the fraction of decaying dark matter  w.r.t. total dark
    matter, $\tau_X$ is the lifetime of dark matter with the corresponding
    decay redshift $z_X$ defined for dark matter with lifetime shorter than
    the age of the Universe. Energy injected in a redshift interval $\Delta
    z$ or timestep, $\Delta t= \frac{|\Delta z|}{(1+z)H(z)}$, is given by,
\begin{equation}
    \Delta E_{\mathrm{inj}}=\frac{f_X}{\tau_X} \rho_c c^2 (1+z)^3 e^{-t/\tau_X}\times\Delta t.
 \label{dmeq1}  
   \end{equation}
    We define the energy deposition fraction at a particular redshift $z$ to
    be the ratio of amount of energy deposited at that redshift, $\Delta
    E_{\rm dep}$, to the energy injected at $t=\tau_X$ within a timestep, i.e. energy deposition fraction $f_{\mathrm{total}}(z)$ is given by,
\begin{equation}
f_{\mathrm{total}}(z)=\frac{\Delta E_{\mathrm{dep}}(z)/\Delta t}{\Delta E_{\mathrm{inj}}/\Delta t}=\frac{\Delta E_{\mathrm{dep}}(z)/\Delta t}{ \frac{f_X}{\tau_X} \rho_c c^2 (1+z)^3 e^{-1.0}},
\label{dmeq2}
\end{equation}   
and similarly for fractions of energy deposited into any particular channel
such as ionization, excitation etc.
    We have checked that our timesteps are small enough {such that}  the energy deposition fraction is independent of the timestep used. The dark matter annihilation module in \textbf{Recfast++} \cite{Seager:1999bc,Chluba2010} uses on-the-spot energy deposition fraction as was suggested in \cite{CK2003}. We have modified this module to use our energy deposition fractions.  \par
    \hspace{1cm}
    \begin{figure}
\centering
\includegraphics[scale=1.0]{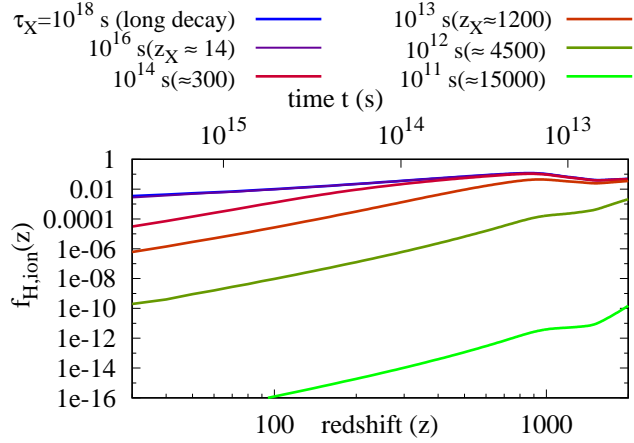}
\caption{Fraction of energy going to direct hydrogen ionization as a function
  of redshift for decay to 100 GeV electron-positron pair for different
  lifetimes. Note that the energy injection histories for
  $\tau_X=10^{18}~{\rm s}$ and $10^{16}~{\rm s}$ are
  identical.}
\label{fig:Hdepfrac}
\end{figure}  
 \begin{figure}
\centering
\includegraphics[scale=1.0]{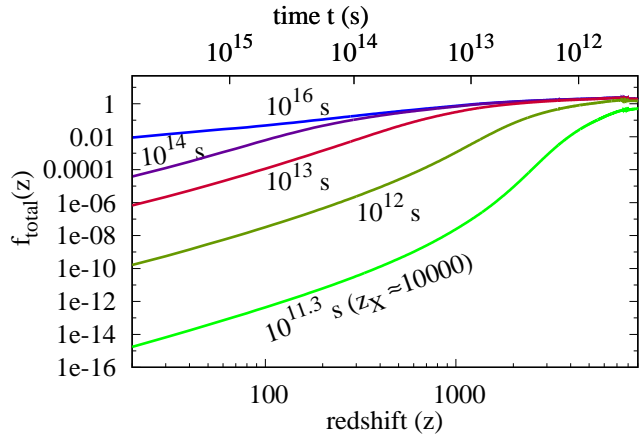}
\caption{{Total energy deposition fraction, as defined in Eq. \ref{dmeq2},}  as a function of redshift for
  decay to 100 GeV electron-positron pair for different
  lifetimes. }
\label{fig:totaldepfrac}
\end{figure} 
We plot the fraction of energy going into hydrogen ionization as a function
of redshift with standard recombination history in
Fig. \ref{fig:Hdepfrac}. We neglect the fact that the recombination history
is modified due to energy injection. This is a good approximation since
slight modification to the recombination history with energy injection will
have a negligible effect on energy deposition fraction. Significant
modification to recombination history is excluded by existing cosmological
data \cite{P20151}. {For long lifetimes ($\tau_X>>10^{13}$s), energy
deposition history ($f_{\mathrm{total}}(z), f_{\rm H,ion}(z)$ etc.) becomes
independent of the lifetime as shown in Fig. \ref{fig:Hdepfrac} and
\ref{fig:totaldepfrac}. This can also be seen from Eq. \ref{dmeq1}. For
$t/\tau_X\rightarrow 0$, the only time dependence that remains is that of
 dilution of dark matter energy density due to the expansion of the Universe which is independent
of the lifetime of the dark matter and the only effect
the lifetime $\tau_X$ has is to change the total amount of injected energy.} For long lifetimes, Eq. \ref{dmeq2} becomes,
\begin{equation}
\frac{\Delta E_{\mathrm{dep}}(z)}{\Delta t}= f_{\mathrm{total}}(z)\frac{f_X}{\tau_X} \rho_c c^2 (1+z)^3 e^{-1.0}.
\end{equation}
Since $f_{\mathrm{total}}(z)$ is independent of $\tau_X$, the deposited
energy is just a function of $f_X/\tau_X$, i.e. $\tau_X$ just affects the
total energy deposited but not how that energy is deposited as a function
of redshift. Therefore, for $\tau_X>>10^{13}$s, the constraints on $f_X$
are defined by constant total energy injection,
i.e. $\frac{f_X}{\tau_X}$=constant or $f_X \propto \tau_X$. We thus need to
derive constraint for only one $\tau_X$ and analytically scale the
constraints for any other $\tau_X$ as long as the condition
$\tau_X>>10^{13}$s is satisfied. For shorter lifetimes, the contribution of
$e^{(-t/\tau_X)}$ {in Eq. \ref{dmeq1}} is important. For $\tau_X\lesssim 10^{13}$ s, most of the energy is deposited in the ionized Universe while some photons survive carrying a small fraction of energy until recombination ($z<$2000). Thus, the constraints become weaker again with decreasing $\tau_X$. Even for $z_X$ as high as 10000 ($\tau_X\approx 10^{11.3}$), a tiny fraction of energy survives until recombination. This can be seen in Fig. \ref{fig:Hdepfrac} as well as in Fig. \ref{fig:totaldepfrac}. This small fraction of surviving energy is still large enough for CMB anisotropies to provide constraints competitive with the other probes. 
\section{\label{sec:calc}Planck CMB constraints on dark matter decay}
\begin{figure}
\centering
\resizebox{\hsize}{!}{\includegraphics{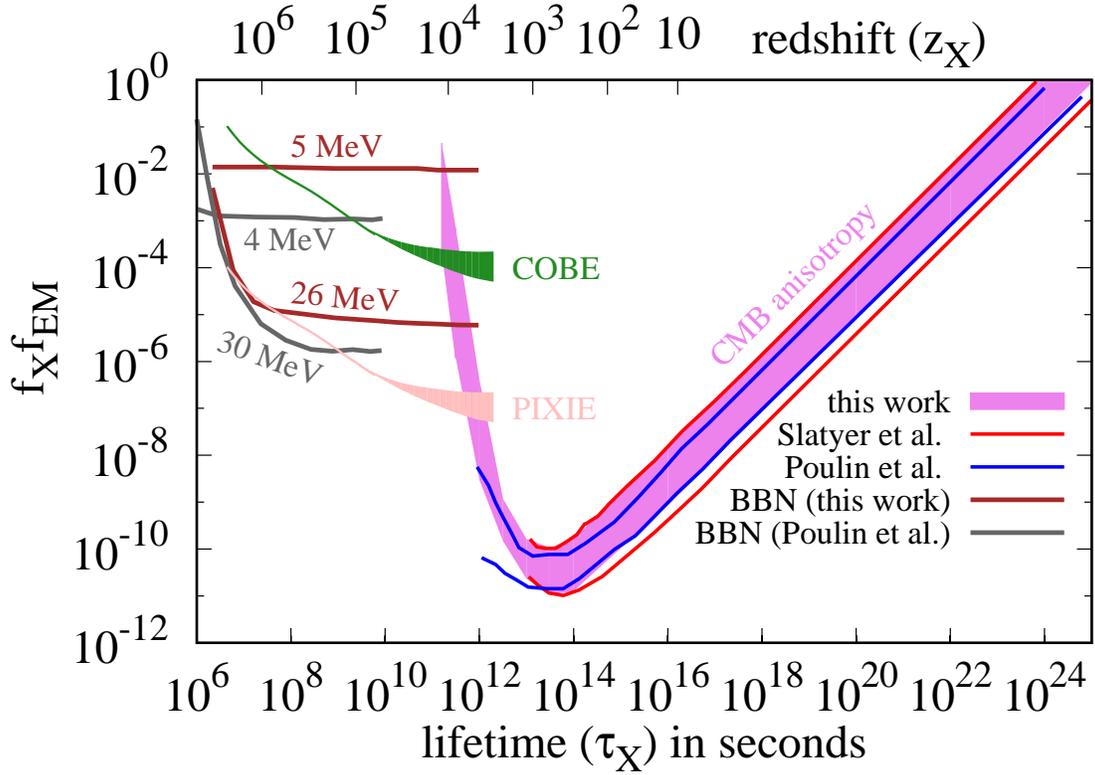}}
\caption{{2-$\sigma$ constraints on electromagnetic decay of dark matter with the
  energy of injected electrons, positrons and photons in the range 10 keV -
  1 TeV. Comparison with previous calculations on CMB anisotropy ( Slatyer
et al. \cite{SC2017}, Poulin et al. \cite{PLS2017}) is also
shown. Energy-dependent spectral distortion constraints of \cite{AK2019}
and extrapolated constraints for Primordial Inflation Explorer (PIXIE) \cite{Pixie2011}, assuming a factor of $1000$
improvement over COBE-FIRAS,  are also shown. The two lines
for BBN are our  strongest constraints from $^4{\rm He}$
destruction (stronger) and $^2{\rm H}$ destruction for photon injection
with the injected  photon
energy of $E_{\gamma}=26~{\rm MeV}$ and $5~{\rm MeV}$ respectively. For comparison we
also show the constraints from Poulin et al. \cite{PS2015} for $E_{\gamma}=
30~{\rm MeV}$  and $E_{\gamma}=4~{\rm MeV}$.}}
\label{fig:result}
\end{figure}

We consider electromagnetic energy injection from dark matter decay to
monochromatic electron-positron pair or photon pair to derive constraints
on fraction of decaying dark matter (compared to total dark matter) from
CMB anisotropy calculations. We do Markov chain Monte Carlo (MCMC) analysis
using publicly available code \textbf{CosmoMC} \cite{LB2002} in combination
with \textbf{Recfast++} module of \textbf{CosmoRec}
\cite{Seager:1999bc,Chluba2010} to get CMB anisotropy constraints as a
function of lifetime and mass of dark matter. We simultaneously fit for
$\Lambda$CDM cosmological parameters plus $f_X$ using Planck 2015
PlikTT,TE,EE and lowTEB likelihood \cite{Plik2015}. Our results are shown
as 2$\sigma$ upper limits {on $f_Xf_{\rm EM}$ in Fig. \ref{fig:result} as a function of
$\tau_X$, where
  $f_{\rm EM}$ is the fraction of  decay energy going into visible
electromagnetic particles and $1-f_{\rm EM}$ is the fraction lost to invisible particles
such as neutrinos, dark radiation or other particles in the dark sector}. CMB anisotropy constraints depend on the initial energy of
injected particle or equivalently the mass of dark matter particle and
varying the mass gives the band in Fig. \ref{fig:result}. The constraints
are strongest for $\tau_X\approx 10^{13}$s ($z_X\approx$ 1200) which is
close to the peak of the CMB visibility function
\cite{Zks1969,Peebles1969}. For both longer and shorter lifetimes (compared
to the  recombination epoch), the constraints get weaker. Our calculations
are in broad agreement for $\tau_x>10^{13}$s with the work of
\cite{SC2017}. For very long lifetime, the constraints follow the scaling
$f_X\propto \tau$ {as discussed in the previous section.} For
lifetime less than $10^{13}$s, the authors in \cite{PLS2017} have used the
energy deposition fraction of \cite{Slatyer20162} to obtain
constraints. However, they have assumed explicit factorization of
deposition fraction to an ionization fraction and a redshift dependent part
which may be invalid when ionization fraction is very high (see Fig. 8 of
\cite{Slatyer2013} and related discussions around Fig. 11 of
\cite{SC2017}). 

{In deriving these constraints we assumed
  monochromatic decay of dark matter, i.e. the energy of injected photons
  $E_{\gamma} = m_X/2$, where $m_X$ is the mass of the decaying dark matter
  particle and the kinetic energy of the injected electron-positron pairs is given by $E_{\rm
    e}= (m_X-m_{\rm e})/2$. The band represents a range of energies from
  $10 ~{\rm keV}$ to $1 ~{\rm TeV}$ for the initial photons, electrons and positrons. Any general decay channel of an
  unstable particle would produce a spectrum of energies for the initial
  photons, electrons and positrons, once the initial unstable standard model
  particles have decayed. The constraints on decay into any channel would 
   still be given by the  band shown in Fig. \ref{fig:result}, as long as
most of the decay energy is in the particles with energy $\gtrsim 10~{\rm
  keV}$.}
\par
\begin{figure}
\centering
\includegraphics[scale=1.0]{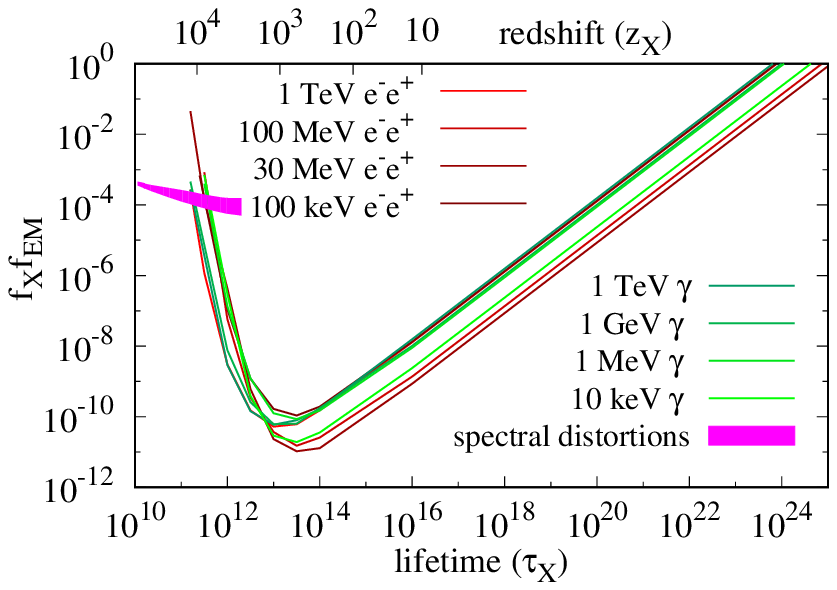}
\caption{Constraints (2-$\sigma$) on electromagnetic decay of  dark matter for different dark matter
  masses (or initial energies of injected particles). Also shown are the
  COBE-FIRAS spectral distortion constraints of \cite{AK2019}.}
\label{fig:result1}
\end{figure}

\hspace{1cm}
In Fig. \ref{fig:result1}, we plot the constraints from dark matter decay
to electron-positron and photon pairs for a few dark matter masses
individually {with $f_{\rm EM}=1$ and assuming monochromatic decay channel.} For $\tau_X>10^{13}$s, constraints for the photon channel are
slightly weaker compared to the electron-positron channel for the same dark
matter mass. The high energy photons are less efficient in depositing their
energies for long lifetimes. {Electrons and positrons with the same energy would produce lower energy}
photons from ICS which are more efficient in depositing their energies to
the  baryonic gas. However, this difference vanishes for very high energies ($>$
GeV). This is to be expected as cascade evolution proceeds through cyclic
production of electron, positrons and photons, making $e^-e^+$  and photon
pair injection indistinguishable. The high energy photons and $e^-e^+$
($\sim$ TeV) give the strongest constraints for short lifetimes
($\tau_X<10^{13}$s) as the high energy gamma rays have a greater chance to
survive until $z<$1000 and deposit a non-negligible fraction of their
energy after recombination thus influencing the residual free electron
fraction. 

 \par
\hspace{1cm} 
We find that the CMB anisotropy constraints are stronger  compared to BBN
and CMB spectral distortions for $z_X\lesssim$ 10000 for all initial
energies except for photon injection just above {the helium
photo-dissociation threshold of $\approx 20~{\rm MeV}$.
The strongest constraints from BBN correspond to injection of photons with
energy $26~{\rm MeV}$ and are shown in Fig. \ref{fig:result}. In this case the BBN
constraints are stronger until significantly lower redshifts ($z\sim
7000-8000$) as explained in next section.  We discuss
the BBN constraints in detail 
in the next section.}


We note that the energy injection from dark matter decay does not have
significant degeneracies with other cosmological parameters
\cite{SC2017}. In particular the best-fit/mean values of the 6 
$\Lambda$CDM parameters does not change significantly. We give the full
parameter table as well as 2-parameter probability density function (PDF) contour plots of dark matter decay
fraction vs standard $\Lambda$CDM parameters for a few illustrative cases in Appendix \ref{app:deg}.

\color{black}
\subsection{CMB spectral distortions} 
 \begin{figure}[!tbp]
  \begin{subfigure}[b]{0.4\textwidth}
    \includegraphics[scale=0.8]{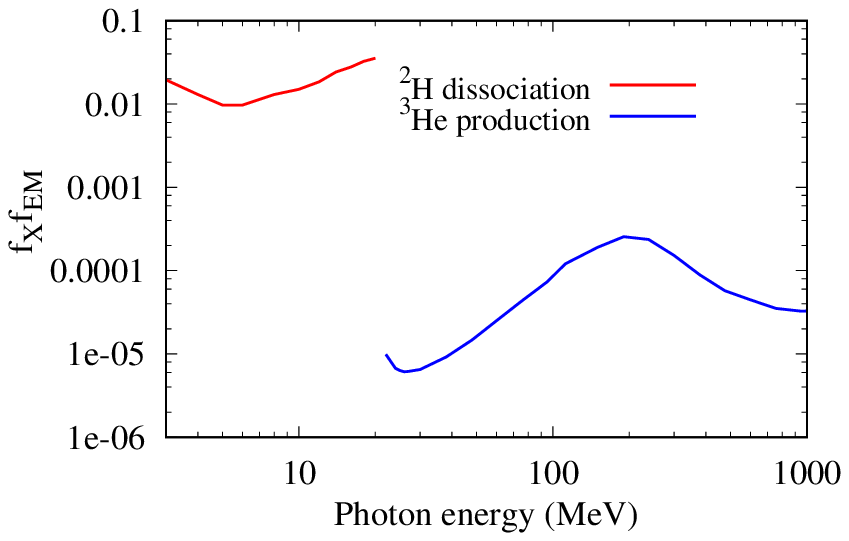}
    \caption{BBN constraints ($2-\sigma$) as a function of injected photon energy for $z_X$=10000.}
    \label{deu}
  \end{subfigure}\hspace{50 pt}
  \begin{subfigure}[b]{0.4\textwidth}
    \includegraphics[scale=0.8]{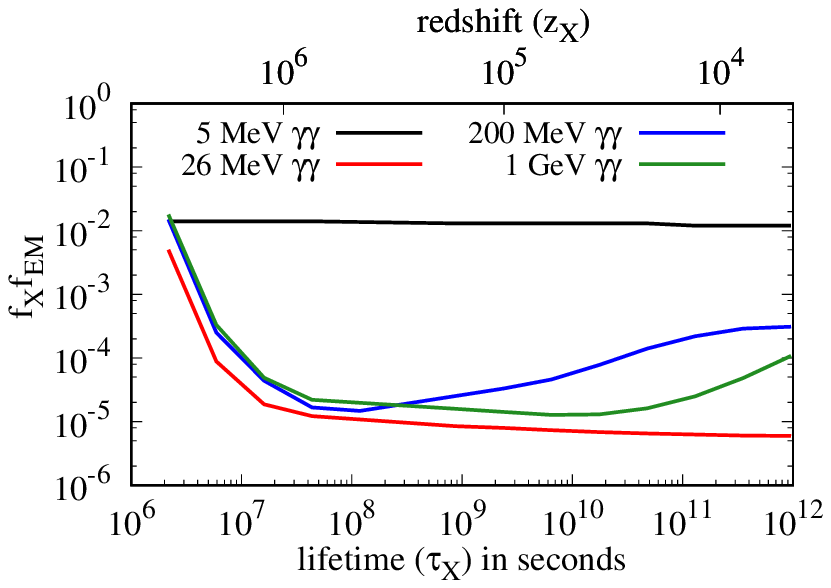}
    \caption{BBN constraints ($2-\sigma$) as a function of lifetime for different
      injected photon energies. }
    \label{He}
  \end{subfigure}
  \caption{Variation of BBN constraints as a function of photon energy and
    lifetime. }
  \label{fig:bbnconst}
\end{figure}
Electromagnetic energy injections at $z\lesssim 2\times 10^6$ result in
distortion of CMB spectrum
from a Planck spectrum \cite{sz1970,dd1982,ks2012,cs2012}. The Cosmic
Background Explorer-Far Infrared Absolute Spectrophotometer (COBE-FIRAS) \cite{Cobe1994} observations of the CMB
spectrum, therefore, constrain energy injection from dark matter decay to
electromagnetic particles. Almost all previous calculations have assumed
the distortions to be y-type ($z \lesssim 10^5$) or $\mu$ type ($10^5
\lesssim z \lesssim 2\times 10^6$) \cite{Sz1969,Is19752}. Recently, we
showed that there assumptions are not strictly correct. In particular, even
after taking into account intermediate or i-type distortions
\cite{Ks2013,Chluba:2013vsa}, e.g. in \cite{AK2019,LSHLC2019}, there is significant corrections to
non-relativistic y-type and i-type distortions \cite{AK2018,AK2019} {since
the injected particles have energy $\gtrsim m_{\rm e}$.}  It was shown in
\cite{AK2019} that, the spectral distortion constraints are energy
dependent and can be relaxed with respect to the constraints obtained
assuming y-type spectrum by a factor of 4 to 5. We use the results of
\cite{AK2019} for spectral distortions constraints in
Fig. \ref{fig:result}. The non-relativistic dark matter energy density
decreases due to the expansion of the Universe as $\propto (1+z)^3$ while
CMB energy density $\rho_{\gamma}\propto (1+z)^4$. The
fractional injected energy density $\frac{\Delta E_{\mathrm{inj}}}{\rho_{\gamma}} \propto \frac{1}{1+z}$ for same fraction $f_X$ of decaying dark matter.  Therefore, for longer decay lifetimes (or lower decay redshifts) constraints on decaying dark matter get stronger in general. However, the scaling is not exactly $(1+z)$ since the electromagnetic cascade evolution also has a redshift dependence \cite{AK2019}.
\par
\hspace{1cm}
 
\begin{table}[h!]
  \begin{center}
    \begin{tabular}{l|c|r} 
      Reactions & photo-dissociation threshold (MeV)  \\
      \hline
      $^2{\rm H}$+$\gamma$ $\rightarrow$ n+p & 2.22 \\
      $^3{\rm He}$+$\gamma$ $\rightarrow$ $^2{\rm H}$+p & 5.49 \\
      $^3{\rm He}$+$\gamma$ $\rightarrow$ n+p+p & 7.718\\
       $^4{\rm He}$+$\gamma$ $\rightarrow$ $^3{\rm H}$+p, $^3{\rm H} \rightarrow^3{\rm He}+e^- +\nu_e$ & 19.81\\
       $^4{\rm He}$+$\gamma$ $\rightarrow$ $^3{\rm He}$+n & 20.58\\
       $^4{\rm He}$+$\gamma$ $\rightarrow$ $^2{\rm H}$+$^2{\rm H}$ & 23.85\\
       $^4{\rm He}$+$\gamma$ $\rightarrow$ $^2{\rm H}$+n+p & 26.07\\
    \end{tabular}
  \end{center}
   \caption{Photo-dissociation reactions included  in this paper and  their threshold energy. }
   \label{tab:table1}
\end{table}
\begin{table}[h!]
\begin{center}
    \begin{tabular}{l|c|r} 
      Elements & theoretical value(1$\sigma$)  & observational value(1$\sigma$)  \\
      \hline
      $n_{^2{\rm H}}/n_{\rm H}$ & $(2.58\overset{+}{-}0.13)\times 10^{-5}$ \cite{CFOY2016} & $(2.53\overset{+}{-}0.04)\times 10^{-5}$  \cite{CFOY2016}\\
     $Y_p$ & $0.24709\overset{+}{-}0.00025$ \cite{CFOY2016} & $0.2449\overset{+}{-}0.0040$ \cite{AOS2015}\\
      $n_{^3{\rm He}}/n_{\rm H}$ & $(10.039\overset{+}{-}0.090)\times 10^{-6}$ \cite{CFOY2016} & $1.5\times 10^{-5}$ (2$\sigma$ upper limit) \cite{BRB2002}\\
    \end{tabular}
  \end{center}
  \caption{Theoretical predictions and observational  bounds on primordial abundance
    of light elements. The ratio of number density of deuterium ($n_{^2{\rm
        H}}$) and helium-3 ($n_{^3{\rm He}}$)  to hydrogen ($n_{\rm H}$) is
    quoted. For Helium-4 ($^4{\rm He}$) the mass fraction, $Y_P$, is the ratio of helium mass density to the total mass density of hydrogen and helium.}
   \label{tab:table2}
\end{table}

\section{BBN constraints}\label{sec:bbn}
High energy photons (energy greater than deuterium ($^2{\rm H}$), helium-3
($^3{\rm He}$) and helium-4 ($^4{\rm He}$) {photo-dissociation}
threshold) can change the abundance of primordial $^2{\rm H}$, $^3{\rm He}$
and $^4{\rm He}$. Note that any $^3{\rm H}$ {produced from destruction of
$^4{\rm He}$} immediately ({on time scale much shorter than the age
of the Universe at recombination}) decays to $^3{\rm He}$. Requiring these
elements to be not over- or under-produced (so that the theoretical
calculations and observations are not in tension), puts constraints on the
energy injection of particles with energy greater than 2.22 MeV. The
 photo-dissociation processes {that are important for changing the
   primordial element abundances and included  in our calculations are given in
   Table \ref{tab:table1} along with the corresponding photo-dissociation thresholds.}
 We use the fits provided in \cite{CEFO2003} {for the
   photo-dissociation cross sections. Abundance of primordial elements can
   constrain electromagnetic energy injection after the primordial
   nucleosynthesis is over at $z\lesssim 10^8$.   For a particular energy
   injection model, the highest redshift  at
   which energy injection can be constrained by BBN depends upon the
   injected photon energy \cite{PS2015}} and its pair-production threshold
 on CMB photons. Since, the CMB photons outnumber baryons by more than nine order of
 magnitude, pair production on CMB dominates every other electromagnetic
 and nuclear process whenever it is kinematically allowed. Pair-production
 on CMB photons produces $e^-e^+$ pairs which produce lower energy photons
 through ICS. {When pair-production threshold is equal to the
 photo-dissociation threshold of a particular reaction at some redshift
 $z_{\rm th}$, all of the energy of the injected photons gets immediately degraded to low energy
 photons that are not able to dissociate nuclei, making BBN constraint from that
 particular reaction exponentially weaker for $z\gtrsim z_{\rm th}$.} 

{The BBN constraints on
   decaying dark   matter have been previously studied in 
   \cite{KM1995,KKMT2018,PS2015,HSW2018,FMW2019}. Early calculations of the effect of electromagnetic energy
  injection on abundances of elements assumed that any high energy particle
  injection results in an initial  \emph{universal photon spectrum},
  independent of the injected particle energy. This assumption was  shown
  to be inaccurate in 
 \cite{PS2015,FMW2019}  for photon injection in the energy range
 $\sim$MeV-$\sim$100 MeV. In particular, it was shown by \cite{PS2015}  that
 the  constraints for photon injection with $\sim$30 MeV energy can be an order
 of magnitude stronger compared to those obtained under the assumption of  \emph{universal photon spectrum} 
 \cite{KM1995}. However, these calculations 
 \cite{KM1995,KKMT2018,PS2015,HSW2018,FMW2019} still  make a quasi-static or  on-the-spot
 approximation to derive the BBN constraints i.e. they assume that energy
 deposition happens on a time scale much faster than the Hubble time. }

{We relax the quasi-static/one-the-spot  assumption also  and  explicitly evolve the electromagnetic
 cascade in an expanding Universe} with our code to obtain the BBN constraints. The theoretical
 predictions and observed abundances for $^2{\rm H}$, $^3{\rm He}$ and
 $^4{\rm He}$ are given in Table \ref{tab:table2}. We use the difference
 between the observed 2-$\sigma$ upper limit and theoretically predicted
 2-$\sigma$ lower limit as the maximum amount of element that can be
 produced by dissociation of a heavier element. Similarly, the difference
 between the theoretical 2-$\sigma$ upper limit and 2-$\sigma$ observed lower
 limit gives the maximum amount of destruction allowed for an element.  {We don't consider $^3{\rm He}$ destruction as there is only
 an upper limit on its abundance. Since the abundance of deuterium is small
 compared to the background electrons and ions by a factor of $\sim 10^{5}$, $^2{\rm H}$ is an unlikely
 target for an energetic photon making the constraint in this case very
 weak. The strongest constraints come from destruction of
$^4{\rm He}$ since it has a much larger abundance, $\approx 8\%$ of hydrogen
by number and the deuterium destruction constraints are  relevant
only below the $^4{\rm He}$ destruction threshold, $E_{\gamma}<19.81~{\rm
 MeV}$. The  photons with energy $\gtrsim 19.81 ~{\rm MeV}$  photo-dissociate 
 $^4{\rm He}$ to $^3{\rm He},^3{\rm H}$ giving strong constraints
 due to  $^3{\rm He}$ over-production.}

{ We show, in
   Figs. \ref{deu} and \ref{He}, the BBN constraints from dark matter decay
   into photons for $z_X$=10000, as a function of photon energy}. {We see
that the constraints depend very strongly on the energy of injected
particles and can vary by more than an order of magnitude. The constraints
are the strongest at  
$E_{\gamma}\approx 26~{\rm MeV}$ and $5~{\rm MeV}$ for $^4{\rm He}$ and $^2{\rm
  H}$ destruction respectively and these strongest limits are shown in
Fig. \ref{fig:result}. We can interpret these constraints for a
general decay channel after defining $f_{\rm EM}$ to be
approximately the energy
injected in electromagnetic particles above the respective thresholds of $2.22~{\rm
  MeV}$ and $19.81~{\rm MeV}$.}  The 5 MeV photon is below the threshold
 of $^4{\rm He}$ photo-dissociation and therefore these photons can only destroy
 $^2{\rm H}~$. We do not show
the constraints from
 $^2{\rm H}$ over-production since the cross-section for $^4{\rm He}$
 destruction to $^3{\rm He}$ is an order of magnitude larger compared to
 the photo-dissociation of $^4{\rm He}$ to $^2{\rm H}$ and thus {the latter gives
 weaker constraints.}  Since the rates
 for all relevant processes which compete with photo-dissociation depend on
 the number density of targets (i.e. background electrons, ions and
 photons) and thus have the same redshift dependence ($\propto
 (1+z)^3$), the constraints are
 independent of dark matter lifetime  for  $\tau_X\gtrsim 10^{8}$s. However, for $\tau_X\lesssim 10^{8}~{\rm s}$, the threshold energy for
 pair production on CMB photons becomes similar to the photo-dissociation
 threshold  and the constraints become
 sensitive to $\tau_X$. 

We also show comparison with the published results from \cite{PS2015} for
$E_{\gamma}=4~{\rm MeV}$ and $30~{\rm MeV}$. Our constraints are weaker by
a factor of 3 for $^3{\rm He}$ overproduction and almost an order of
magnitude for $^2{\rm H}$ destruction. This is expected, since we evolve
the electromagnetic cascades in the expanding Universe and allow the
energy to be deposited not instantaneously but in a delayed manner. There
is thus a higher probability for the energy to be lost to other processes
such as heating, decreasing the fraction of energy going into the destruction of
primordial elements. 

\section{Conclusions}    
In this work, we obtain constraints on decaying dark matter from Planck
observations of CMB temperature and polarization anisotropy power
spectra and abundances of light elements. Our CMB anisotropy results broadly agree with the previous calculations for dark
matter lifetime $\tau_X\gtrsim 10^{13}$s. We give the first CMB
anisotropies constraints for  $\tau_X< 10^{13}$s upto the point where the
constraints from CMB spectral distortions and BBN are stronger. We find
that for general energies, the CMB anisotropies are a powerful probe of
energy injection for redshifts as high as 10000, providing the strongest
constraints today and thereby excluding a new region of parameter
space. For these calculations, we have developed a new code which evolves
high energy particle cascades taking into account background evolution of
ionization fraction of hydrogen, helium, singly ionized helium and
background expansion. We do not rely on factorization of the full
calculation into low and high energy part and evolve both sub-keV and
higher energy particles consistently in one unified code. {For completeness,
we also show the CMB spectral distortion constraints from COBE-FIRAS
data and give forecasts for a future PIXIE like mission.} 
We also calculate
primordial elements abundance constraints with the same code and find that
taking delayed deposition of energy is important for accurate
constraints. Taking the delayed deposition of energy into account weakens
the constraints considerably compared to the instantaneous deposition
approximation made in previous calculations.  

Even though, we
have only considered monochromatic electron-positron and photon pair
injections, it is straightforward to interpret our constraints for any
general decay channel. Our calculations  can also be generalized to other
energy injection processes with arbitrary spectrum, which we leave for
future work. One such example would
be  evaporation
of primordial black holes.  Our calculations suggest that the CMB anisotropies  can constrain black hole
evaporation at  higher redshifts or lower black hole masses than what has
been considered in the literature so far.    
\par
\hspace{1cm}
 Although the CMB anisotropy analysis only provides constraints on amount
 of energy injection, our calculations can be extended to a more direct
 probe for electromagnetic energy injection around or before recombination,
 namely the cosmological recombination spectrum. Since, recombination is a
 non-equilibrium process, a characteristic distortion signal with
 information about recombination as well as background CMB spectrum gets
 imprinted on the initially Planckian CMB spectrum. Energy injection around
 recombination modifies this signal in an unique way \cite{C2010}. We have
 argued that  the excitations to higher atomic energy levels will be important for accurate predictions of recombination line spectral
 distortions. We however leave a detailed study for future work.

 \section{Acknowledgements}
 We acknowledge the use of computational facilities of Department of
 Theoretical Physics at Tata Institute of Fundamental Research,
 Mumbai. This work was supported by Max Planck Partner Group for cosmology of Max Planck
 Institute for Astrophysics Garching at 
 Tata Institute of Fundamental Research funded by
 Max-Planck-Gesellschaft. This work was also supported by 
Science and Engineering Research Board (SERB) of Department of Science and
Technology, Government of India grant no. ECR/2015/000078. We acknowledge support of the Department of Atomic Energy, Government of India, under project no. 12-R\&D-TFR-5.02-0200.

\bibliographystyle{unsrtads}
\bibliography{cmbconst}

\begin{thebibliography}{10}

\bibitem{Pl2018}
N.~Aghanim et~al.
\newblock {Planck 2018 results. VI. Cosmological parameters}.
\newblock {\em ArXiv e-prints}, July 2018.
\newblock \href {http://arxiv.org/abs/1807.06209} {\path{arXiv:1807.06209}},
  {\small[\href{http://adsabs.harvard.edu/abs/2018arXiv180706209P}{ADS}]}.

\bibitem{SS2000}
David~N. {Spergel} and Paul~J. {Steinhardt}.
\newblock {Observational Evidence for Self-Interacting Cold Dark Matter}.
\newblock {\em \prl}, 84(17):3760--3763, Apr 2000.
\newblock \href {http://arxiv.org/abs/astro-ph/9909386}
  {\path{arXiv:astro-ph/9909386}}, \href
  {http://dx.doi.org/10.1103/PhysRevLett.84.3760} {\path{[DOI]}},
  {\small[\href{https://ui.adsabs.harvard.edu/abs/2000PhRvL..84.3760S}{ADS}]}.

\bibitem{BKS2017}
Torsten {Bringmann}, Felix {Kahlhoefer}, Kai {Schmidt-Hoberg}, and Parampreet
  {Walia}.
\newblock {Strong Constraints on Self-Interacting Dark Matter with Light
  Mediators}.
\newblock {\em \prl}, 118(14):141802, Apr 2017.
\newblock \href {http://arxiv.org/abs/1612.00845} {\path{arXiv:1612.00845}},
  \href {http://dx.doi.org/10.1103/PhysRevLett.118.141802} {\path{[DOI]}},
  {\small[\href{https://ui.adsabs.harvard.edu/abs/2017PhRvL.118n1802B}{ADS}]}.

\bibitem{BMS2015}
Manuel~A. {Buen-Abad}, Gustavo {Marques-Tavares}, and Martin {Schmaltz}.
\newblock {Non-Abelian dark matter and dark radiation}.
\newblock {\em \prd}, 92(2):023531, Jul 2015.
\newblock \href {http://arxiv.org/abs/1505.03542} {\path{arXiv:1505.03542}},
  \href {http://dx.doi.org/10.1103/PhysRevD.92.023531} {\path{[DOI]}},
  {\small[\href{https://ui.adsabs.harvard.edu/abs/2015PhRvD..92b3531B}{ADS}]}.

\bibitem{LMS2016}
Julien {Lesgourgues}, Gustavo {Marques-Tavares}, and Martin {Schmaltz}.
\newblock {Evidence for dark matter interactions in cosmological precision
  data?}
\newblock {\em \jcap}, 2016(2):037, Feb 2016.
\newblock \href {http://arxiv.org/abs/1507.04351} {\path{arXiv:1507.04351}},
  \href {http://dx.doi.org/10.1088/1475-7516/2016/02/037} {\path{[DOI]}},
  {\small[\href{https://ui.adsabs.harvard.edu/abs/2016JCAP...02..037L}{ADS}]}.

\bibitem{ASS1998}
J.~A. {Adams}, S.~{Sarkar}, and D.~W. {Sciama}.
\newblock {Cosmic microwave background anisotropy in the decaying neutrino
  cosmology}.
\newblock {\em \mnras}, 301:210--214, November 1998.
\newblock \href {http://arxiv.org/abs/astro-ph/9805108}
  {\path{arXiv:astro-ph/9805108}}, \href
  {http://dx.doi.org/10.1046/j.1365-8711.1998.02017.x} {\path{[DOI]}},
  {\small[\href{http://adsabs.harvard.edu/abs/1998MNRAS.301..210A}{ADS}]}.

\bibitem{CK2003}
X.~{Chen} and M.~{Kamionkowski}.
\newblock {Particle decays during the cosmic dark ages}.
\newblock {\em \prd}, 70(4):043502, August 2004.
\newblock \href {http://arxiv.org/abs/astro-ph/0310473}
  {\path{arXiv:astro-ph/0310473}}, \href
  {http://dx.doi.org/10.1103/PhysRevD.70.043502} {\path{[DOI]}},
  {\small[\href{http://adsabs.harvard.edu/abs/2004PhRvD..70d3502C}{ADS}]}.

\bibitem{H1975}
S.~W. {Hawking}.
\newblock {Particle creation by black holes}.
\newblock {\em Communications in Mathematical Physics}, 43:199--220, August
  1975.
\newblock \href {http://dx.doi.org/10.1007/BF02345020} {\path{[DOI]}},
  {\small[\href{https://ui.adsabs.harvard.edu/abs/1975CMaPh..43..199H}{ADS}]}.

\bibitem{SKLP2018}
Patrick {St{\"o}cker}, Michael {Kr{\"a}mer}, Julien {Lesgourgues}, and Vivian
  {Poulin}.
\newblock {Exotic energy injection with ExoCLASS: application to the Higgs
  portal model and evaporating black holes}.
\newblock {\em \jcap}, 2018(3):018, Mar 2018.
\newblock \href {http://arxiv.org/abs/1801.01871} {\path{arXiv:1801.01871}},
  \href {http://dx.doi.org/10.1088/1475-7516/2018/03/018} {\path{[DOI]}},
  {\small[\href{https://ui.adsabs.harvard.edu/abs/2018JCAP...03..018S}{ADS}]}.

\bibitem{Slatyer:2009yq}
T.~R. {Slatyer}, N.~{Padmanabhan}, and D.~P. {Finkbeiner}.
\newblock {CMB constraints on WIMP annihilation: Energy absorption during the
  recombination epoch}.
\newblock {\em \prd}, 80(4):043526, August 2009.
\newblock \href {http://arxiv.org/abs/0906.1197} {\path{arXiv:0906.1197}},
  \href {http://dx.doi.org/10.1103/PhysRevD.80.043526} {\path{[DOI]}},
  {\small[\href{http://adsabs.harvard.edu/abs/2009PhRvD..80d3526S}{ADS}]}.

\bibitem{hutsi}
G.~{H{\"u}tsi}, J.~{Chluba}, A.~{Hektor}, and M.~{Raidal}.
\newblock {WMAP7 and future CMB constraints on annihilating dark matter:
  implications for GeV-scale WIMPs}.
\newblock {\em \aap}, 535:A26, Nov 2011.
\newblock \href {http://arxiv.org/abs/1103.2766} {\path{arXiv:1103.2766}},
  \href {http://dx.doi.org/10.1051/0004-6361/201116914} {\path{[DOI]}},
  {\small[\href{https://ui.adsabs.harvard.edu/abs/2011A&A...535A..26H}{ADS}]}.

\bibitem{Slatyer20161}
Tracy~R. {Slatyer}.
\newblock {Indirect dark matter signatures in the cosmic dark ages. I.
  Generalizing the bound on s -wave dark matter annihilation from Planck
  results}.
\newblock {\em \prd}, 93(2):023527, Jan 2016.
\newblock \href {http://arxiv.org/abs/1506.03811} {\path{arXiv:1506.03811}},
  \href {http://dx.doi.org/10.1103/PhysRevD.93.023527} {\path{[DOI]}},
  {\small[\href{https://ui.adsabs.harvard.edu/abs/2016PhRvD..93b3527S}{ADS}]}.

\bibitem{DBK2014}
Cora {Dvorkin}, Kfir {Blum}, and Marc {Kamionkowski}.
\newblock {Constraining dark matter-baryon scattering with linear cosmology}.
\newblock {\em \prd}, 89(2):023519, Jan 2014.
\newblock \href {http://arxiv.org/abs/1311.2937} {\path{arXiv:1311.2937}},
  \href {http://dx.doi.org/10.1103/PhysRevD.89.023519} {\path{[DOI]}},
  {\small[\href{https://ui.adsabs.harvard.edu/abs/2014PhRvD..89b3519D}{ADS}]}.

\bibitem{ALM2014}
Benjamin {Audren}, Julien {Lesgourgues}, Gianpiero {Mangano}, Pasquale~Dario
  {Serpico}, and Thomas {Tram}.
\newblock {Strongest model-independent bound on the lifetime of Dark Matter}.
\newblock {\em \jcap}, 2014(12):028, Dec 2014.
\newblock \href {http://arxiv.org/abs/1407.2418} {\path{arXiv:1407.2418}},
  \href {http://dx.doi.org/10.1088/1475-7516/2014/12/028} {\path{[DOI]}},
  {\small[\href{https://ui.adsabs.harvard.edu/abs/2014JCAP...12..028A}{ADS}]}.

\bibitem{W1982}
S.~{Weinberg}.
\newblock {Cosmological constraints on the scale of supersymmetry breaking}.
\newblock {\em Physical Review Letters}, 48:1303--1306, May 1982.
\newblock \href {http://dx.doi.org/10.1103/PhysRevLett.48.1303} {\path{[DOI]}},
  {\small[\href{https://ui.adsabs.harvard.edu/abs/1982PhRvL..48.1303W}{ADS}]}.

\bibitem{G1983}
H.~{Goldberg}.
\newblock {Constraint on the photino mass from cosmology}.
\newblock {\em Physical Review Letters}, 50:1419--1422, May 1983.
\newblock \href {http://dx.doi.org/10.1103/PhysRevLett.50.1419} {\path{[DOI]}},
  {\small[\href{https://ui.adsabs.harvard.edu/abs/1983PhRvL..50.1419G}{ADS}]}.

\bibitem{MMY1993}
T.~{Moroi}, H.~{Murayama}, and M.~{Yamaguchi}.
\newblock {Cosmological constraints on the light stable gravitino}.
\newblock {\em Physics Letters B}, 303:289--294, April 1993.
\newblock \href {http://dx.doi.org/10.1016/0370-2693(93)91434-O}
  {\path{[DOI]}},
  {\small[\href{https://ui.adsabs.harvard.edu/abs/1993PhLB..303..289M}{ADS}]}.

\bibitem{FRT2003}
Jonathan~L. {Feng}, Arvind {Rajaraman}, and Fumihiro {Takayama}.
\newblock {Superweakly Interacting Massive Particles}.
\newblock {\em \prl}, 91:011302, Jul 2003.
\newblock \href {http://arxiv.org/abs/hep-ph/0302215}
  {\path{arXiv:hep-ph/0302215}}, \href
  {http://dx.doi.org/10.1103/PhysRevLett.91.011302} {\path{[DOI]}},
  {\small[\href{https://ui.adsabs.harvard.edu/\#abs/2003PhRvL..91a1302F}{ADS}]}.

\bibitem{FRT20031}
Jonathan~L. {Feng}, Arvind {Rajaraman}, and Fumihiro {Takayama}.
\newblock {Superweakly interacting massive particle dark matter signals from
  the early Universe}.
\newblock {\em \prd}, 68:063504, Sep 2003.
\newblock \href {http://arxiv.org/abs/hep-ph/0306024}
  {\path{arXiv:hep-ph/0306024}}, \href
  {http://dx.doi.org/10.1103/PhysRevD.68.063504} {\path{[DOI]}},
  {\small[\href{https://ui.adsabs.harvard.edu/\#abs/2003PhRvD..68f3504F}{ADS}]}.

\bibitem{CFM2002}
Hsin-Chia {Cheng}, Jonathan~L. {Feng}, and Konstantin~T. {Matchev}.
\newblock {Kaluza-Klein Dark Matter}.
\newblock {\em \prl}, 89:211301, Oct 2002.
\newblock \href {http://arxiv.org/abs/hep-ph/0207125}
  {\path{arXiv:hep-ph/0207125}}, \href
  {http://dx.doi.org/10.1103/PhysRevLett.89.211301} {\path{[DOI]}},
  {\small[\href{https://ui.adsabs.harvard.edu/\#abs/2002PhRvL..89u1301C}{ADS}]}.

\bibitem{AFSW2009}
Nima {Arkani-Hamed}, Douglas~P. {Finkbeiner}, Tracy~R. {Slatyer}, and Neal
  {Weiner}.
\newblock {A theory of dark matter}.
\newblock {\em \prd}, 79(1):015014, Jan 2009.
\newblock \href {http://arxiv.org/abs/0810.0713} {\path{arXiv:0810.0713}},
  \href {http://dx.doi.org/10.1103/PhysRevD.79.015014} {\path{[DOI]}},
  {\small[\href{https://ui.adsabs.harvard.edu/abs/2009PhRvD..79a5014A}{ADS}]}.

\bibitem{BW2012}
Torsten Bringmann and Christoph Weniger.
\newblock Gamma ray signals from dark matter: Concepts, status and prospects.
\newblock {\em Physics of the Dark Universe}, 1(1):194 -- 217, 2012.
\newblock Next Decade in Dark Matter and Dark Energy.
\newblock URL:
  \url{http://www.sciencedirect.com/science/article/pii/S221268641200009X},
  \href {http://dx.doi.org/https://doi.org/10.1016/j.dark.2012.10.005}
  {\path{[DOI]}}.

\bibitem{F2010}
Jonathan~L. {Feng}.
\newblock {Dark Matter Candidates from Particle Physics and Methods of
  Detection}.
\newblock {\em Annual Review of Astronomy and Astrophysics}, 48:495--545, Sep
  2010.
\newblock \href {http://arxiv.org/abs/1003.0904} {\path{arXiv:1003.0904}},
  \href {http://dx.doi.org/10.1146/annurev-astro-082708-101659} {\path{[DOI]}},
  {\small[\href{https://ui.adsabs.harvard.edu/\#abs/2010ARA&A..48..495F}{ADS}]}.

\bibitem{BHS2005}
Gianfranco {Bertone}, Dan {Hooper}, and Joseph {Silk}.
\newblock {Particle dark matter: evidence, candidates and constraints}.
\newblock {\em \physrep}, 405(5-6):279--390, Jan 2005.
\newblock \href {http://arxiv.org/abs/hep-ph/0404175}
  {\path{arXiv:hep-ph/0404175}}, \href
  {http://dx.doi.org/10.1016/j.physrep.2004.08.031} {\path{[DOI]}},
  {\small[\href{https://ui.adsabs.harvard.edu/abs/2005PhR...405..279B}{ADS}]}.

\bibitem{Galli:2013dna}
S.~{Galli}, T.~R. {Slatyer}, M.~{Valdes}, and F.~{Iocco}.
\newblock {Systematic uncertainties in constraining dark matter annihilation
  from the cosmic microwave background}.
\newblock {\em \prd}, 88(6):063502, September 2013.
\newblock \href {http://arxiv.org/abs/1306.0563} {\path{arXiv:1306.0563}},
  \href {http://dx.doi.org/10.1103/PhysRevD.88.063502} {\path{[DOI]}},
  {\small[\href{http://adsabs.harvard.edu/abs/2013PhRvD..88f3502G}{ADS}]}.

\bibitem{Zks1969}
Y.~B. {Zeldovich}, V.~G. {Kurt}, and R.~A. {Sunyaev}.
\newblock {Recombination of Hydrogen in the Hot Model of the Universe}.
\newblock {\em Soviet Journal of Experimental and Theoretical Physics}, 28:146,
  January 1969.
\newblock
  {\small[\href{http://adsabs.harvard.edu/abs/1969JETP...28..146Z}{ADS}]}.

\bibitem{Peebles1969}
P.~J.~E. {Peebles}.
\newblock {Recombination of the Primeval Plasma}.
\newblock {\em \apj}, 153:1, July 1968.
\newblock \href {http://dx.doi.org/10.1086/149628} {\path{[DOI]}},
  {\small[\href{http://adsabs.harvard.edu/abs/1968ApJ...153....1P}{ADS}]}.

\bibitem{Padmanabhan:2005es}
N.~{Padmanabhan} and D.~P. {Finkbeiner}.
\newblock {Detecting dark matter annihilation with CMB polarization: Signatures
  and experimental prospects}.
\newblock {\em \prd}, 72(2):023508, July 2005.
\newblock \href {http://arxiv.org/abs/astro-ph/0503486}
  {\path{arXiv:astro-ph/0503486}}, \href
  {http://dx.doi.org/10.1103/PhysRevD.72.023508} {\path{[DOI]}},
  {\small[\href{http://adsabs.harvard.edu/abs/2005PhRvD..72b3508P}{ADS}]}.

\bibitem{Galli:2009zc}
S.~{Galli}, F.~{Iocco}, G.~{Bertone}, and A.~{Melchiorri}.
\newblock {CMB constraints on dark matter models with large annihilation cross
  section}.
\newblock {\em \prd}, 80(2):023505, July 2009.
\newblock \href {http://arxiv.org/abs/0905.0003} {\path{arXiv:0905.0003}},
  \href {http://dx.doi.org/10.1103/PhysRevD.80.023505} {\path{[DOI]}},
  {\small[\href{http://adsabs.harvard.edu/abs/2009PhRvD..80b3505G}{ADS}]}.

\bibitem{SC2017}
Tracy~R. {Slatyer} and Chih-Liang {Wu}.
\newblock {General constraints on dark matter decay from the cosmic microwave
  background}.
\newblock {\em \prd}, 95(2):023010, Jan 2017.
\newblock \href {http://arxiv.org/abs/1610.06933} {\path{arXiv:1610.06933}},
  \href {http://dx.doi.org/10.1103/PhysRevD.95.023010} {\path{[DOI]}},
  {\small[\href{https://ui.adsabs.harvard.edu/abs/2017PhRvD..95b3010S}{ADS}]}.

\bibitem{PLS2017}
Vivian {Poulin}, Julien {Lesgourgues}, and Pasquale~D. {Serpico}.
\newblock {Cosmological constraints on exotic injection of electromagnetic
  energy}.
\newblock {\em \jcap}, 2017(3):043, Mar 2017.
\newblock \href {http://arxiv.org/abs/1610.10051} {\path{arXiv:1610.10051}},
  \href {http://dx.doi.org/10.1088/1475-7516/2017/03/043} {\path{[DOI]}},
  {\small[\href{https://ui.adsabs.harvard.edu/abs/2017JCAP...03..043P}{ADS}]}.

\bibitem{SV1985}
J.~M. {Shull} and M.~E. {van Steenberg}.
\newblock {X-ray secondary heating and ionization in quasar emission-line
  clouds}.
\newblock {\em \apj}, 298:268--274, November 1985.
\newblock \href {http://dx.doi.org/10.1086/163605} {\path{[DOI]}},
  {\small[\href{https://ui.adsabs.harvard.edu/abs/1985ApJ...298..268S}{ADS}]}.

\bibitem{FS2010}
Steven~R. {Furlanetto} and Samuel~Johnson {Stoever}.
\newblock {Secondary ionization and heating by fast electrons}.
\newblock {\em \mnras}, 404(4):1869--1878, Jun 2010.
\newblock \href {http://arxiv.org/abs/0910.4410} {\path{arXiv:0910.4410}},
  \href {http://dx.doi.org/10.1111/j.1365-2966.2010.16401.x} {\path{[DOI]}},
  {\small[\href{https://ui.adsabs.harvard.edu/abs/2010MNRAS.404.1869F}{ADS}]}.

\bibitem{VEF2010}
M.~{Vald{\'e}s}, C.~{Evoli}, and A.~{Ferrara}.
\newblock {Particle energy cascade in the intergalactic medium}.
\newblock {\em \mnras}, 404(3):1569--1582, May 2010.
\newblock \href {http://arxiv.org/abs/0911.1125} {\path{arXiv:0911.1125}},
  \href {http://dx.doi.org/10.1111/j.1365-2966.2010.16387.x} {\path{[DOI]}},
  {\small[\href{https://ui.adsabs.harvard.edu/abs/2010MNRAS.404.1569V}{ADS}]}.

\bibitem{KK2008}
Toru {Kanzaki} and Masahiro {Kawasaki}.
\newblock {Electron and photon energy deposition in the Universe}.
\newblock {\em \prd}, 78(10):103004, Nov 2008.
\newblock \href {http://arxiv.org/abs/0805.3969} {\path{arXiv:0805.3969}},
  \href {http://dx.doi.org/10.1103/PhysRevD.78.103004} {\path{[DOI]}},
  {\small[\href{https://ui.adsabs.harvard.edu/abs/2008PhRvD..78j3004K}{ADS}]}.

\bibitem{KKN2010}
T.~{Kanzaki}, M.~{Kawasaki}, and K.~{Nakayama}.
\newblock {Effects of Dark Matter Annihilation on the Cosmic Microwave
  Background}.
\newblock {\em Progress of Theoretical Physics}, 123(5):853--865, May 2010.
\newblock \href {http://arxiv.org/abs/0907.3985} {\path{arXiv:0907.3985}},
  \href {http://dx.doi.org/10.1143/PTP.123.853} {\path{[DOI]}},
  {\small[\href{https://ui.adsabs.harvard.edu/abs/2010PThPh.123..853K}{ADS}]}.

\bibitem{P20151}
{Planck Collaboration} and P.~A.~R. {Ade}.
\newblock {Planck 2015 results. XIII. Cosmological parameters}.
\newblock {\em \aap}, 594:A13, Sep 2016.
\newblock \href {http://arxiv.org/abs/1502.01589} {\path{arXiv:1502.01589}},
  \href {http://dx.doi.org/10.1051/0004-6361/201525830} {\path{[DOI]}},
  {\small[\href{https://ui.adsabs.harvard.edu/abs/2016A&A...594A..13P}{ADS}]}.

\bibitem{Slatyer20162}
Tracy~R. {Slatyer}.
\newblock {Indirect dark matter signatures in the cosmic dark ages. II.
  Ionization, heating, and photon production from arbitrary energy injections}.
\newblock {\em \prd}, 93(2):023521, Jan 2016.
\newblock \href {http://arxiv.org/abs/1506.03812} {\path{arXiv:1506.03812}},
  \href {http://dx.doi.org/10.1103/PhysRevD.93.023521} {\path{[DOI]}},
  {\small[\href{https://ui.adsabs.harvard.edu/abs/2016PhRvD..93b3521S}{ADS}]}.

\bibitem{ENS1985}
J.~{Ellis}, D.~V. {Nanopoulos}, and S.~{Sarkar}.
\newblock {The cosmology of decaying gravitinos}.
\newblock {\em Nuclear Physics B}, 259:175--188, September 1985.
\newblock \href {http://dx.doi.org/10.1016/0550-3213(85)90306-2}
  {\path{[DOI]}},
  {\small[\href{https://ui.adsabs.harvard.edu/abs/1985NuPhB.259..175E}{ADS}]}.

\bibitem{EGLNS1992}
J.~{Ellis}, G.~B. {Gelmini}, J.~L. {Lopez}, D.~V. {Nanopoulos}, and
  S.~{Sarkar}.
\newblock {Astrophysical constraints on massive unstable neutral relic
  particles}.
\newblock {\em Nuclear Physics B}, 373:399--437, April 1992.
\newblock \href {http://dx.doi.org/10.1016/0550-3213(92)90438-H}
  {\path{[DOI]}},
  {\small[\href{https://ui.adsabs.harvard.edu/abs/1992NuPhB.373..399E}{ADS}]}.

\bibitem{KM1995}
M.~{Kawasaki} and T.~{Moroi}.
\newblock {Electromagnetic Cascade in the Early Universe and Its Application to
  the Big Bang Nucleosynthesis}.
\newblock {\em \apj}, 452:506, Oct 1995.
\newblock \href {http://arxiv.org/abs/astro-ph/9412055}
  {\path{arXiv:astro-ph/9412055}}, \href {http://dx.doi.org/10.1086/176324}
  {\path{[DOI]}},
  {\small[\href{https://ui.adsabs.harvard.edu/abs/1995ApJ...452..506K}{ADS}]}.

\bibitem{KKMT2018}
Masahiro {Kawasaki}, Kazunori {Kohri}, Takeo {Moroi}, and Yoshitaro {Takaesu}.
\newblock {Revisiting big-bang nucleosynthesis constraints on long-lived
  decaying particles}.
\newblock {\em \prd}, 97(2):023502, Jan 2018.
\newblock \href {http://arxiv.org/abs/1709.01211} {\path{arXiv:1709.01211}},
  \href {http://dx.doi.org/10.1103/PhysRevD.97.023502} {\path{[DOI]}},
  {\small[\href{https://ui.adsabs.harvard.edu/abs/2018PhRvD..97b3502K}{ADS}]}.

\bibitem{PS2015}
Vivian {Poulin} and Pasquale~Dario {Serpico}.
\newblock {Nonuniversal BBN bounds on electromagnetically decaying particles}.
\newblock {\em \prd}, 91(10):103007, May 2015.
\newblock \href {http://arxiv.org/abs/1503.04852} {\path{arXiv:1503.04852}},
  \href {http://dx.doi.org/10.1103/PhysRevD.91.103007} {\path{[DOI]}},
  {\small[\href{https://ui.adsabs.harvard.edu/abs/2015PhRvD..91j3007P}{ADS}]}.

\bibitem{HSW2018}
Marco {Hufnagel}, Kai {Schmidt-Hoberg}, and Sebastian {Wild}.
\newblock {BBN constraints on MeV-scale dark sectors. Part II: Electromagnetic
  decays}.
\newblock {\em \jcap}, 2018(11):032, Nov 2018.
\newblock \href {http://arxiv.org/abs/1808.09324} {\path{arXiv:1808.09324}},
  \href {http://dx.doi.org/10.1088/1475-7516/2018/11/032} {\path{[DOI]}},
  {\small[\href{https://ui.adsabs.harvard.edu/abs/2018JCAP...11..032H}{ADS}]}.

\bibitem{FMW2019}
Lindsay {Forestell}, David~E. {Morrissey}, and Graham {White}.
\newblock {Limits from BBN on light electromagnetic decays}.
\newblock {\em Journal of High Energy Physics}, 2019(1):74, Jan 2019.
\newblock \href {http://arxiv.org/abs/1809.01179} {\path{arXiv:1809.01179}},
  \href {http://dx.doi.org/10.1007/JHEP01(2019)074} {\path{[DOI]}},
  {\small[\href{https://ui.adsabs.harvard.edu/abs/2019JHEP...01..074F}{ADS}]}.

\bibitem{AK2018}
Sandeep~Kumar {Acharya} and Rishi {Khatri}.
\newblock {Rich structure of nonthermal relativistic CMB spectral distortions
  from high energy particle cascades at redshifts $z\lesssim 2 {\times}
  10^{5}$}.
\newblock {\em \prd}, 99(4):043520, Feb 2019.
\newblock \href {http://arxiv.org/abs/1808.02897} {\path{arXiv:1808.02897}},
  \href {http://dx.doi.org/10.1103/PhysRevD.99.043520} {\path{[DOI]}},
  {\small[\href{https://ui.adsabs.harvard.edu/abs/2019PhRvD..99d3520A}{ADS}]}.

\bibitem{Chluba2010}
J.~{Chluba} and R.~M. {Thomas}.
\newblock {Towards a complete treatment of the cosmological recombination
  problem}.
\newblock {\em \mnras}, 412:748--764, April 2011.
\newblock \href {http://arxiv.org/abs/1010.3631} {\path{arXiv:1010.3631}},
  \href {http://dx.doi.org/10.1111/j.1365-2966.2010.17940.x} {\path{[DOI]}},
  {\small[\href{http://adsabs.harvard.edu/abs/2011MNRAS.412..748C}{ADS}]}.

\bibitem{Seager:1999bc}
S.~{Seager}, D.~D. {Sasselov}, and D.~{Scott}.
\newblock {A New Calculation of the Recombination Epoch}.
\newblock {\em \apjl}, 523:L1--L5, September 1999.
\newblock \href {http://arxiv.org/abs/astro-ph/9909275}
  {\path{arXiv:astro-ph/9909275}}, \href {http://dx.doi.org/10.1086/312250}
  {\path{[DOI]}},
  {\small[\href{http://adsabs.harvard.edu/abs/1999ApJ...523L...1S}{ADS}]}.

\bibitem{seager2000}
Sara {Seager}, Dimitar~D. {Sasselov}, and Douglas {Scott}.
\newblock {How Exactly Did the Universe Become Neutral?}
\newblock {\em \apjs}, 128(2):407--430, Jun 2000.
\newblock \href {http://arxiv.org/abs/astro-ph/9912182}
  {\path{arXiv:astro-ph/9912182}}, \href {http://dx.doi.org/10.1086/313388}
  {\path{[DOI]}},
  {\small[\href{https://ui.adsabs.harvard.edu/abs/2000ApJS..128..407S}{ADS}]}.

\bibitem{CS2006}
J.~{Chluba} and R.~A. {Sunyaev}.
\newblock {Induced two-photon decay of the 2s level and the rate of
  cosmological hydrogen recombination}.
\newblock {\em \aap}, 446:39--42, January 2006.
\newblock \href {http://arxiv.org/abs/astro-ph/0508144}
  {\path{arXiv:astro-ph/0508144}}, \href
  {http://dx.doi.org/10.1051/0004-6361:20053988} {\path{[DOI]}},
  {\small[\href{https://ui.adsabs.harvard.edu/abs/2006A%26A...446...39C}{ADS}]}.

\bibitem{SH2008}
Eric~R. {Switzer} and Christopher~M. {Hirata}.
\newblock {Primordial helium recombination. I. Feedback, line transfer, and
  continuum opacity}.
\newblock {\em \prd}, 77(8):083006, Apr 2008.
\newblock \href {http://arxiv.org/abs/astro-ph/0702143}
  {\path{arXiv:astro-ph/0702143}}, \href
  {http://dx.doi.org/10.1103/PhysRevD.77.083006} {\path{[DOI]}},
  {\small[\href{https://ui.adsabs.harvard.edu/abs/2008PhRvD..77h3006S}{ADS}]}.

\bibitem{RCS2008}
J.~A. {Rubi{\~n}o-Mart{\'\i}n}, J.~{Chluba}, and R.~A. {Sunyaev}.
\newblock {Lines in the cosmic microwave background spectrum from the epoch of
  cosmological helium recombination}.
\newblock {\em \aap}, 485(2):377--393, Jul 2008.
\newblock \href {http://arxiv.org/abs/0711.0594} {\path{arXiv:0711.0594}},
  \href {http://dx.doi.org/10.1051/0004-6361:20078993} {\path{[DOI]}},
  {\small[\href{https://ui.adsabs.harvard.edu/abs/2008A&A...485..377R}{ADS}]}.

\bibitem{GH2010}
Daniel {Grin} and Christopher~M. {Hirata}.
\newblock {Cosmological hydrogen recombination: The effect of extremely high-n
  states}.
\newblock {\em \prd}, 81(8):083005, Apr 2010.
\newblock \href {http://arxiv.org/abs/0911.1359} {\path{arXiv:0911.1359}},
  \href {http://dx.doi.org/10.1103/PhysRevD.81.083005} {\path{[DOI]}},
  {\small[\href{https://ui.adsabs.harvard.edu/abs/2010PhRvD..81h3005G}{ADS}]}.

\bibitem{CVD2010}
J.~{Chluba}, G.~M. {Vasil}, and L.~J. {Dursi}.
\newblock {Recombinations to the Rydberg states of hydrogen and their effect
  during the cosmological recombination epoch}.
\newblock {\em \mnras}, 407(1):599--612, Sep 2010.
\newblock \href {http://arxiv.org/abs/1003.4928} {\path{arXiv:1003.4928}},
  \href {http://dx.doi.org/10.1111/j.1365-2966.2010.16940.x} {\path{[DOI]}},
  {\small[\href{https://ui.adsabs.harvard.edu/abs/2010MNRAS.407..599C}{ADS}]}.

\bibitem{AH2010}
Yacine {Ali-Ha{\"\i}moud} and Christopher~M. {Hirata}.
\newblock {Ultrafast effective multilevel atom method for primordial hydrogen
  recombination}.
\newblock {\em \prd}, 82(6):063521, Sep 2010.
\newblock \href {http://arxiv.org/abs/1006.1355} {\path{arXiv:1006.1355}},
  \href {http://dx.doi.org/10.1103/PhysRevD.82.063521} {\path{[DOI]}},
  {\small[\href{https://ui.adsabs.harvard.edu/abs/2010PhRvD..82f3521A}{ADS}]}.

\bibitem{Hh2011}
Y.~{Ali-Ha{\"i}moud} and C.~M. {Hirata}.
\newblock {HyRec: A fast and highly accurate primordial hydrogen and helium
  recombination code}.
\newblock {\em \prd}, 83(4):043513, February 2011.
\newblock \href {http://arxiv.org/abs/1011.3758} {\path{arXiv:1011.3758}},
  \href {http://dx.doi.org/10.1103/PhysRevD.83.043513} {\path{[DOI]}},
  {\small[\href{http://adsabs.harvard.edu/abs/2011PhRvD..83d3513A}{ADS}]}.

\bibitem{C2010}
J.~{Chluba}.
\newblock {Could the cosmological recombination spectrum help us understand
  annihilating dark matter?}
\newblock {\em \mnras}, 402(2):1195--1207, Feb 2010.
\newblock \href {http://arxiv.org/abs/0910.3663} {\path{arXiv:0910.3663}},
  \href {http://dx.doi.org/10.1111/j.1365-2966.2009.15957.x} {\path{[DOI]}},
  {\small[\href{https://ui.adsabs.harvard.edu/abs/2010MNRAS.402.1195C}{ADS}]}.

\bibitem{LRS2019}
Hongwan {Liu}, Gregory~W. {Ridgway}, and Tracy~R. {Slatyer}.
\newblock {DarkHistory: A code package for calculating modified cosmic
  ionization and thermal histories with dark matter and other exotic energy
  injections}.
\newblock {\em arXiv e-prints}, page arXiv:1904.09296, Apr 2019.
\newblock \href {http://arxiv.org/abs/1904.09296} {\path{arXiv:1904.09296}},
  {\small[\href{https://ui.adsabs.harvard.edu/abs/2019arXiv190409296L}{ADS}]}.

\bibitem{CA2016}
Jens {Chluba} and Yacine {Ali-Ha{\"\i}moud}.
\newblock {COSMOSPEC: fast and detailed computation of the cosmological
  recombination radiation from hydrogen and helium}.
\newblock {\em \mnras}, 456(4):3494--3508, Mar 2016.
\newblock \href {http://arxiv.org/abs/1510.03877} {\path{arXiv:1510.03877}},
  \href {http://dx.doi.org/10.1093/mnras/stv2691} {\path{[DOI]}},
  {\small[\href{https://ui.adsabs.harvard.edu/abs/2016MNRAS.456.3494C}{ADS}]}.

\bibitem{LB2002}
Antony {Lewis} and Sarah {Bridle}.
\newblock {Cosmological parameters from CMB and other data: A Monte Carlo
  approach}.
\newblock {\em \prd}, 66(10):103511, Nov 2002.
\newblock \href {http://arxiv.org/abs/astro-ph/0205436}
  {\path{arXiv:astro-ph/0205436}}, \href
  {http://dx.doi.org/10.1103/PhysRevD.66.103511} {\path{[DOI]}},
  {\small[\href{https://ui.adsabs.harvard.edu/abs/2002PhRvD..66j3511L}{ADS}]}.

\bibitem{Plik2015}
Aghanim et.~al. {Planck Collaboration}.
\newblock {Planck 2015 results. XI. CMB power spectra, likelihoods, and
  robustness of parameters}.
\newblock {\em \aap}, 594:A11, Sep 2016.
\newblock \href {http://arxiv.org/abs/1507.02704} {\path{arXiv:1507.02704}},
  \href {http://dx.doi.org/10.1051/0004-6361/201526926} {\path{[DOI]}},
  {\small[\href{https://ui.adsabs.harvard.edu/abs/2016A&A...594A..11P}{ADS}]}.

\bibitem{Slatyer2013}
Tracy~R. {Slatyer}.
\newblock {Energy injection and absorption in the cosmic dark ages}.
\newblock {\em \prd}, 87(12):123513, Jun 2013.
\newblock \href {http://arxiv.org/abs/1211.0283} {\path{arXiv:1211.0283}},
  \href {http://dx.doi.org/10.1103/PhysRevD.87.123513} {\path{[DOI]}},
  {\small[\href{https://ui.adsabs.harvard.edu/abs/2013PhRvD..87l3513S}{ADS}]}.

\bibitem{AK2019}
Sandeep~Kumar {Acharya} and Rishi {Khatri}.
\newblock {New CMB spectral distortion constraints on decaying dark matter with
  full evolution of electromagnetic cascades before recombination}.
\newblock {\em Physical Review D}, 99(12):123510, Jun 2019.
\newblock \href {http://arxiv.org/abs/1903.04503} {\path{arXiv:1903.04503}},
  \href {http://dx.doi.org/10.1103/PhysRevD.99.123510} {\path{[DOI]}},
  {\small[\href{https://ui.adsabs.harvard.edu/abs/2019PhRvD..99l3510A}{ADS}]}.

\bibitem{Pixie2011}
A.~{Kogut}, D.~J. {Fixsen}, D.~T. {Chuss}, J.~{Dotson}, E.~{Dwek},
  M.~{Halpern}, G.~F. {Hinshaw}, S.~M. {Meyer}, S.~H. {Moseley}, M.~D.
  {Seiffert}, D.~N. {Spergel}, and E.~J. {Wollack}.
\newblock {The Primordial Inflation Explorer (PIXIE): a nulling polarimeter for
  cosmic microwave background observations}.
\newblock {\em \jcap}, 2011(7):025, Jul 2011.
\newblock \href {http://arxiv.org/abs/1105.2044} {\path{arXiv:1105.2044}},
  \href {http://dx.doi.org/10.1088/1475-7516/2011/07/025} {\path{[DOI]}},
  {\small[\href{https://ui.adsabs.harvard.edu/abs/2011JCAP...07..025K}{ADS}]}.

\bibitem{sz1970}
R.~A. {Sunyaev} and Ya.~B. {Zeldovich}.
\newblock {The interaction of matter and radiation in the hot model of the
  Universe, II}.
\newblock {\em \apss}, 7(1):20--30, Apr 1970.
\newblock \href {http://dx.doi.org/10.1007/BF00653472} {\path{[DOI]}},
  {\small[\href{https://ui.adsabs.harvard.edu/abs/1970Ap&SS...7...20S}{ADS}]}.

\bibitem{dd1982}
L.~{Danese} and G.~{de Zotti}.
\newblock {Double Compton process and the spectrum of the microwave
  background}.
\newblock {\em \aap}, 107(1):39--42, Mar 1982.
\newblock
  {\small[\href{https://ui.adsabs.harvard.edu/abs/1982A&A...107...39D}{ADS}]}.

\bibitem{ks2012}
Rishi {Khatri} and Rashid~A. {Sunyaev}.
\newblock {Creation of the CMB spectrum: precise analytic solutions for the
  blackbody photosphere}.
\newblock {\em \jcap}, 2012(6):038, Jun 2012.
\newblock \href {http://arxiv.org/abs/1203.2601} {\path{arXiv:1203.2601}},
  \href {http://dx.doi.org/10.1088/1475-7516/2012/06/038} {\path{[DOI]}},
  {\small[\href{https://ui.adsabs.harvard.edu/abs/2012JCAP...06..038K}{ADS}]}.

\bibitem{cs2012}
J.~{Chluba} and R.~A. {Sunyaev}.
\newblock {The evolution of CMB spectral distortions in the early Universe}.
\newblock {\em \mnras}, 419(2):1294--1314, Jan 2012.
\newblock \href {http://arxiv.org/abs/1109.6552} {\path{arXiv:1109.6552}},
  \href {http://dx.doi.org/10.1111/j.1365-2966.2011.19786.x} {\path{[DOI]}},
  {\small[\href{https://ui.adsabs.harvard.edu/abs/2012MNRAS.419.1294C}{ADS}]}.

\bibitem{Cobe1994}
J.~C. {Mather}, E.~S. {Cheng}, D.~A. {Cottingham}, R.~E. {Eplee}, Jr., D.~J.
  {Fixsen}, T.~{Hewagama}, R.~B. {Isaacman}, K.~A. {Jensen}, S.~S. {Meyer},
  P.~D. {Noerdlinger}, S.~M. {Read}, L.~P. {Rosen}, R.~A. {Shafer}, E.~L.
  {Wright}, C.~L. {Bennett}, N.~W. {Boggess}, M.~G. {Hauser}, T.~{Kelsall},
  S.~H. {Moseley}, Jr., R.~F. {Silverberg}, G.~F. {Smoot}, R.~{Weiss}, and
  D.~T. {Wilkinson}.
\newblock {Measurement of the cosmic microwave background spectrum by the COBE
  FIRAS instrument}.
\newblock {\em \apj}, 420:439--444, January 1994.
\newblock \href {http://dx.doi.org/10.1086/173574} {\path{[DOI]}},
  {\small[\href{http://adsabs.harvard.edu/abs/1994ApJ...420..439M}{ADS}]}.

\bibitem{Sz1969}
Y.~B. {Zeldovich} and R.~A. {Sunyaev}.
\newblock {The Interaction of Matter and Radiation in a Hot-Model Universe}.
\newblock {\em \apss}, 4:301--316, July 1969.
\newblock \href {http://dx.doi.org/10.1007/BF00661821} {\path{[DOI]}},
  {\small[\href{http://adsabs.harvard.edu/abs/1969Ap%26SS...4..301Z}{ADS}]}.

\bibitem{Is19752}
A.~F. {Illarionov} and R.~A. {Siuniaev}.
\newblock {Comptonization, the background-radiation spectrum, and the thermal
  history of the universe}.
\newblock {\em \sovast}, 18:691--699, June 1975.
\newblock
  {\small[\href{http://adsabs.harvard.edu/abs/1975SvA....18..691I}{ADS}]}.

\bibitem{Ks2013}
R.~{Khatri} and R.~A. {Sunyaev}.
\newblock {Beyond y and {$\mu$}: the shape of the CMB spectral distortions in
  the intermediate epoch, $1.5 {\times} 10^{4}\lesssim$z $ \lesssim 2 {\times}
  10^{5}$}.
\newblock {\em \jcap}, 9:016, September 2012.
\newblock \href {http://arxiv.org/abs/1207.6654} {\path{arXiv:1207.6654}},
  \href {http://dx.doi.org/10.1088/1475-7516/2012/09/016} {\path{[DOI]}},
  {\small[\href{http://adsabs.harvard.edu/abs/2012JCAP...09..016K}{ADS}]}.

\bibitem{Chluba:2013vsa}
J.~{Chluba}.
\newblock {Green's function of the cosmological thermalization problem}.
\newblock {\em \mnras}, 434:352--357, September 2013.
\newblock \href {http://arxiv.org/abs/1304.6120} {\path{arXiv:1304.6120}},
  \href {http://dx.doi.org/10.1093/mnras/stt1025} {\path{[DOI]}},
  {\small[\href{http://adsabs.harvard.edu/abs/2013MNRAS.434..352C}{ADS}]}.

\bibitem{LSHLC2019}
Matteo {Lucca}, Nils {Sch{\"o}neberg}, Deanna~C. {Hooper}, Julien
  {Lesgourgues}, and Jens {Chluba}.
\newblock {The synergy between CMB spectral distortions and anisotropies}.
\newblock {\em arXiv e-prints}, page arXiv:1910.04619, Oct 2019.
\newblock \href {http://arxiv.org/abs/1910.04619} {\path{arXiv:1910.04619}},
  {\small[\href{https://ui.adsabs.harvard.edu/abs/2019arXiv191004619L}{ADS}]}.

\bibitem{CFOY2016}
Richard~H. {Cyburt}, Brian~D. {Fields}, Keith~A. {Olive}, and Tsung-Han {Yeh}.
\newblock {Big bang nucleosynthesis: Present status}.
\newblock {\em Reviews of Modern Physics}, 88(1):015004, Jan 2016.
\newblock \href {http://arxiv.org/abs/1505.01076} {\path{arXiv:1505.01076}},
  \href {http://dx.doi.org/10.1103/RevModPhys.88.015004} {\path{[DOI]}},
  {\small[\href{https://ui.adsabs.harvard.edu/abs/2016RvMP...88a5004C}{ADS}]}.

\bibitem{AOS2015}
Erik {Aver}, Keith~A. {Olive}, and Evan~D. {Skillman}.
\newblock {The effects of He I {\ensuremath{\lambda}}10830 on helium abundance
  determinations}.
\newblock {\em Journal of Cosmology and Astro-Particle Physics}, 2015(7):011,
  Jul 2015.
\newblock \href {http://arxiv.org/abs/1503.08146} {\path{arXiv:1503.08146}},
  \href {http://dx.doi.org/10.1088/1475-7516/2015/07/011} {\path{[DOI]}},
  {\small[\href{https://ui.adsabs.harvard.edu/abs/2015JCAP...07..011A}{ADS}]}.

\bibitem{BRB2002}
T.~M. {Bania}, Robert~T. {Rood}, and Dana~S. {Balser}.
\newblock {The cosmological density of baryons from observations of 3He+ in the
  Milky Way}.
\newblock {\em Nature}, 415, 2002.
\newblock \href {http://dx.doi.org/10.1038/415054a} {\path{[DOI]}}.

\bibitem{CEFO2003}
Richard~H. {Cyburt}, John {Ellis}, Brian~D. {Fields}, and Keith~A. {Olive}.
\newblock {Updated nucleosynthesis constraints on unstable relic particles}.
\newblock {\em Physical Review D}, 67(10):103521, May 2003.
\newblock \href {http://arxiv.org/abs/astro-ph/0211258}
  {\path{arXiv:astro-ph/0211258}}, \href
  {http://dx.doi.org/10.1103/PhysRevD.67.103521} {\path{[DOI]}},
  {\small[\href{https://ui.adsabs.harvard.edu/abs/2003PhRvD..67j3521C}{ADS}]}.

\bibitem{I1971}
M.~{Inokuti}.
\newblock {Inelastic Collisions of Fast Charged Particles with Atoms and
  Molecules - The Bethe Theory Revisited}.
\newblock {\em Reviews of Modern Physics}, 43:297--347, July 1971.
\newblock \href {http://dx.doi.org/10.1103/RevModPhys.43.297} {\path{[DOI]}},
  {\small[\href{https://ui.adsabs.harvard.edu/abs/1971RvMP...43..297I}{ADS}]}.

\bibitem{IIT1978}
M.~{Inokuti}, Y.~{Itikawa}, and J.~E. {Turner}.
\newblock {Addenda: Inelastic collisions of fast charged particles with atoms
  and molecules - The Bethe theory revisited}.
\newblock {\em Reviews of Modern Physics}, 50:23--35, January 1978.
\newblock \href {http://dx.doi.org/10.1103/RevModPhys.50.23} {\path{[DOI]}},
  {\small[\href{https://ui.adsabs.harvard.edu/abs/1978RvMP...50...23I}{ADS}]}.

\bibitem{SKD2002}
P.~M. {Stone}, Y.~K. {Kim}, and J.~P. {Desclaux}.
\newblock {Electron-Impact Cross Sections for Dipole- and Spin-Allowed
  Excitations of Hydrogen, Helium, and Lithium}.
\newblock {\em Journal of research of the National Institute of Standards and
  Technology, 107(4), 327–337. doi:10.6028/jres.107.026}, 2002.

\bibitem{AR1985}
M.~{Arnaud} and R.~{Rothenflug}.
\newblock {An updated evaluation of recombination and ionization rates}.
\newblock {\em \aaps}, 60:425--457, June 1985.
\newblock
  {\small[\href{https://ui.adsabs.harvard.edu/abs/1985A%26AS...60..425A}{ADS}]}.

\bibitem{Sz1989}
A.~A. {Zdziarski} and R.~{Svensson}.
\newblock {Absorption of X-rays and gamma rays at cosmological distances}.
\newblock {\em \apj}, 344:551--566, September 1989.
\newblock \href {http://dx.doi.org/10.1086/167826} {\path{[DOI]}},
  {\small[\href{http://adsabs.harvard.edu/abs/1989ApJ...344..551Z}{ADS}]}.

\bibitem{Pl2015}
{Planck Collaboration}.
\newblock {Planck 2015 results. XIII. Cosmological parameters}.
\newblock {\em \aap}, 594:A13, Sep 2016.
\newblock \href {http://arxiv.org/abs/1502.01589} {\path{arXiv:1502.01589}},
  \href {http://dx.doi.org/10.1051/0004-6361/201525830} {\path{[DOI]}},
  {\small[\href{https://ui.adsabs.harvard.edu/abs/2016A&A...594A..13P}{ADS}]}.

\end{thebibliography}
\appendix
\section{Collision cross-sections and energy loss rates for electrons and positrons}\label{app:elec}
The cross-sections and energy-loss rates for electron and positron except
for excitation and ionization of neutral hydrogen and helium are given with references in
\cite{AK2018}. We give below the cross sections for the additional atomic
processes included in the present paper.

\subsection{Collisional excitation of neutral Hydrogen and Helium}
For collisional excitation of neutral hydrogen, we use the tabulated cross-section in CCC data base \footnote{\url{http://atom.curtin.edu.au/CCC-WWW/}} for incident electron with kinetic energy from threshold to $\sim$ keV energy. For higher energy, we use Bethe approximation \cite{I1971, IIT1978, KK2008},
\begin{align}
\sigma(E)_{2p}&=\frac{4\pi a^2_0}{(E/13.6)}\left[0.55\times \ln (\frac{4C_{2p}E}{13.6})+\frac{0.21}{(E/13.6)}\right],\\
   \sigma(E)_{2s}&=\frac{4\pi a^2_0}{(E/13.6)}\left[0.12+\frac{-0.31}{(E/13.6)}\right],\\
   \sigma(E)_{n=3}&=\frac{4\pi a^2_0}{(E/13.6)}\left[8.9\times 10^{-2} \ln (\frac{4C_3E}{13.6})\right],
   \end{align}
{where $a_0$ is the Bohr radius, $\ln C_{2p}=-0.9$, and  $\ln C_3=-0.27$.}
   We have only considered 3p level with above cross-section. Cross-section
   for s level is typically {one order of magnitude} less compared to the p level. \par
   \hspace{1cm}
   For neutral helium excitation to 2p level, we use the data and fit provided in \cite{SKD2002},
   \begin{equation}
   \sigma_{eHe}=\frac{4\pi a^2_0 R}{T+B+E}\left[a\ln (T/R)+b+c\frac{R}{T}\right],
   \end{equation}
   {where $R=$Rydberg energy, $T=$ kinetic energy of electron, $E=$ excitation
   energy, $B=$binding energy of the electron to be excited, $a=0.17, b=-0.08,$
   and $c=0.035$.}
   \subsection{Collisional ionization of neutral Hydrogen and Helium}
   For collisional {ionization of} neutral hydrogen at low energy, we use the cross-section tabulated in CCC database. For high energy,  we use bethe approximation \cite{I1971,IIT1978,KK2008},
   \begin{equation}
   \sigma(E)=\frac{4\pi a^2_0}{E/13.6}\left[0.28\times \ln (\frac{4C_i E}{13.6})+\frac{\gamma_i}{(E/13.6)}\right],
   \end{equation}
   where $\ln C_i=3.048,\gamma_i=-1.63+\ln (13.6/E)$.
   The spectrum of secondary electrons is given by
   \begin{equation}
   \frac{d\sigma(E,\epsilon)}{d\epsilon}=\frac{A(E)}{1+(\epsilon/\epsilon_0)^2} 
   \end{equation}
   with $0 \leqslant \epsilon \leqslant 0.5\times (E-I)$, where $I$ is the ionization threshold, $\epsilon_0$= 8 eV, $A(E)=\frac{\sigma(E)}{\epsilon_0}\left[\tan^{-1}\left(X(E)\right)\right]^{-1}$ with $X(E)=\frac{E-I}{2\epsilon_0}$. 
 For neutral helium cross-section we use the fit \cite{AR1985}, \cite{Slatyer:2009yq},
 \begin{equation}
 \sigma(E)= 10^{-14}{\rm cm}^2 \frac{1}{u(I/eV)^2}\left[A(1-\frac{1}{u})+B(1-\frac{1}{u})^2+C\ln u+\frac{D\ln u}{u}\right],
 \end{equation}
 where $u=E/I$, $A=17.8, B=-11, C=7, D= -23.2$, and  $I=24.6$ eV.
 The spectrum of secondary electrons is same as that of hydrogen with $\epsilon_0=15.8$ eV \cite{FS2010}. \par
 \hspace{1cm}
 We neglect collisional excitation and ionization of singly ionized helium
 as during its recombination epoch ($z\sim$ 6000), the Universe is highly
 ionized and sub-keV electrons deposit most of their energy as heat.

\section{Photo-ionization}
\label{app:photon}
The photo-ionization cross-section used in this work is given by \citep{Slatyer:2009yq,Sz1989},
\begin{equation}
\sigma=\frac{2^9\pi^2 {r_0}^2}{3\alpha^3}{\left(\frac{E_{\mathrm{thres}}}{E}\right)}^4\frac{\mathrm{exp}(-4\eta \arctan{(1/\eta)})}{1-\exp{(-2\pi\eta)}},
\end{equation}
where $\eta=\frac{1}{\left((\frac{E}{E_{\mathrm{thres}}})-1\right)^{1/2}}$,
$r_0$ is the electron radius, $E_{\mathrm{thres}}$=13.6 eV for hydrogen and
54.4 eV for singly ionized helium, and $\alpha$ is the fine structure constant. Photo-ionization cross-section of neutral helium is given by, 
\begin{equation}
\sigma_{\rm He}=-12\sigma_H+5.1\times 10^{-20} {\rm cm}^2 (\frac{E}{250 eV})^{-2.65} 
\end{equation}
for 50 eV $<$ E $<$ 250 eV and
\begin{equation}
\sigma_{\rm He}=-12\sigma_H+5.1\times 10^{-20} {\rm cm}^2 (\frac{E}{250 eV})^{-3.3} 
\end{equation}
for E $>$ 250 eV. {For other processes involving high energy photons, we
refer the reader to \cite{AK2018}.}
\section{Effect of dark matter decay on the recombnation history and CMB anisotropies}
\label{app:plots}
\begin{figure}[!tbp]
  \begin{subfigure}[b]{0.4\textwidth}
    \includegraphics[scale=1.0]{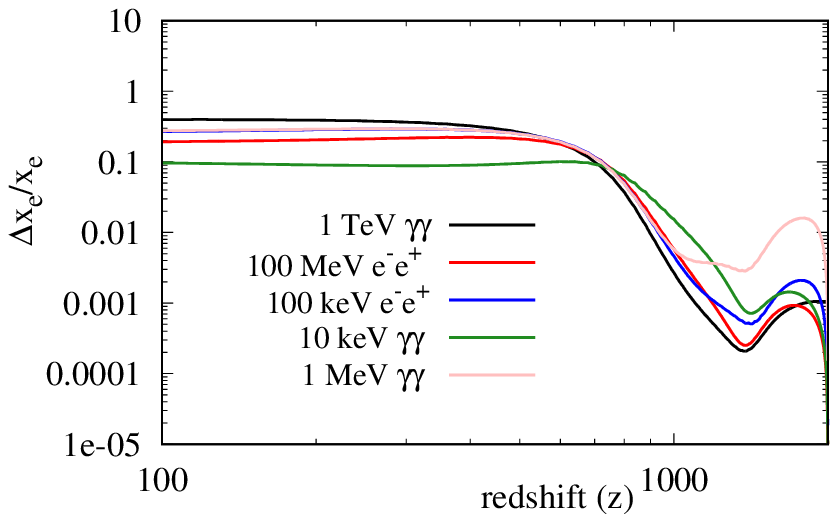}
     \caption{ $\tau_X=10^{12}$s.}
     \label{fig:xetau12}
  \end{subfigure}\hspace{65 pt}
  \begin{subfigure}[b]{0.4\textwidth}
    \includegraphics[scale=1.0]{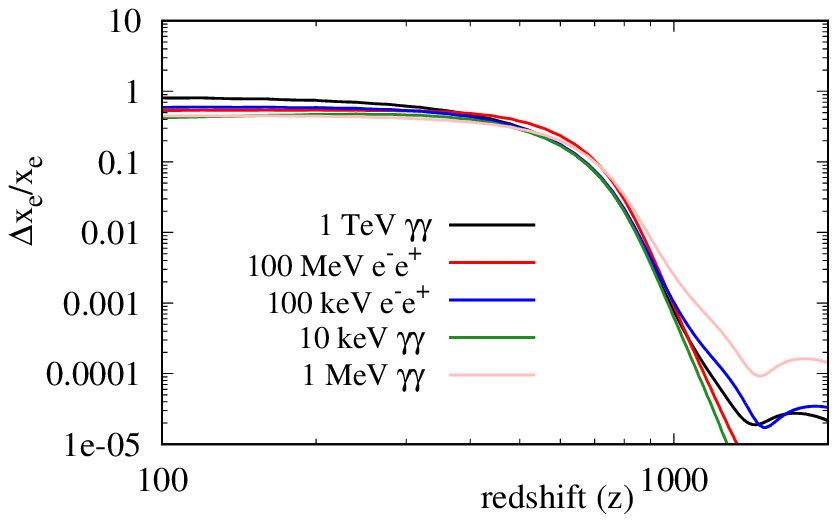}
     \caption{ $\tau_X=10^{13}$s.}
     \label{fig:xetau13}
  \end{subfigure}\hspace{65 pt}
\caption{Change in recombination history for different dark matter decay scenarios}\label{fig:xe}
\end{figure}

\begin{figure}{}
   \begin{subfigure}[b]{0.4\textwidth}
    \includegraphics[scale=1.0]{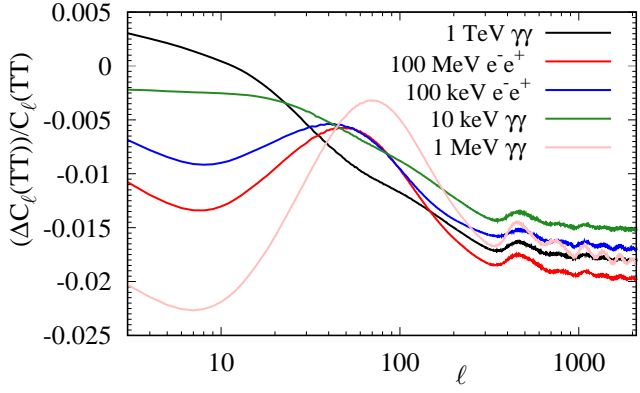}
     \caption{ $\tau_X=10^{13}$s.}
     \label{fig:ClTTtau13}
  \end{subfigure}\hspace{65 pt}
  \begin{subfigure}[b]{0.4\textwidth}
    \includegraphics[scale=1.0]{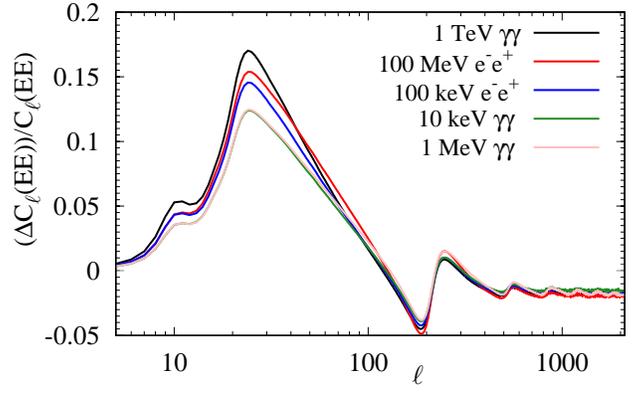}
     \caption{ $\tau_X=10^{13}$s.}
    \label{fig:ClEEtau13}
    \end{subfigure}
    
    \begin{subfigure}[b]{0.4\textwidth}
    \includegraphics[scale=1.0]{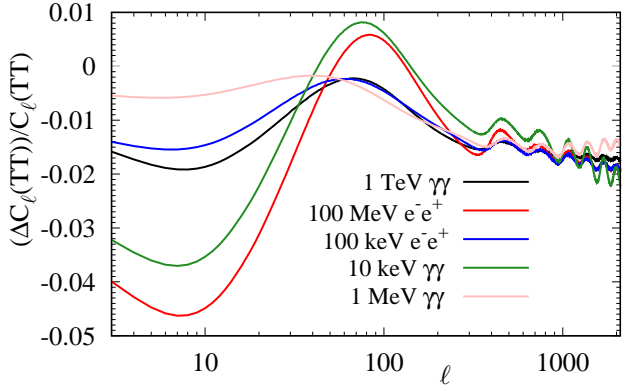}
     \caption{ $\tau_X=10^{12}$s.}
  \label{fig:ClTTtau12}  
  \end{subfigure}\hspace{65 pt}
  \begin{subfigure}[b]{0.4\textwidth}
    \includegraphics[scale=1.0]{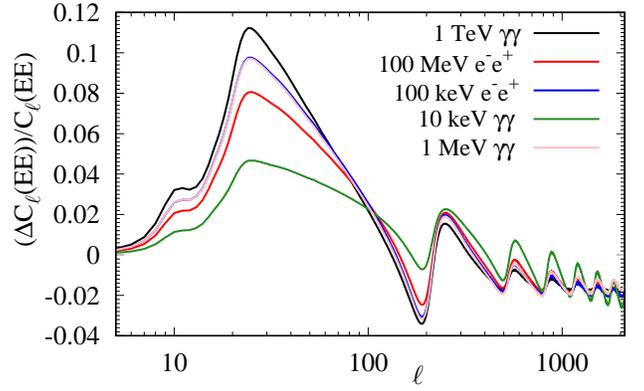}
     \caption{ $\tau_X=10^{12}$s}
   \label{fig:ClEEtau12}   
   \end{subfigure}
 \caption{Fractional change in CMB temperature amd polarizarion power spectrum for
     different dark matter decay scenarios.}
   \label{fig:ps}
   \end{figure}

\begin{figure}
  \includegraphics[scale=1.0]{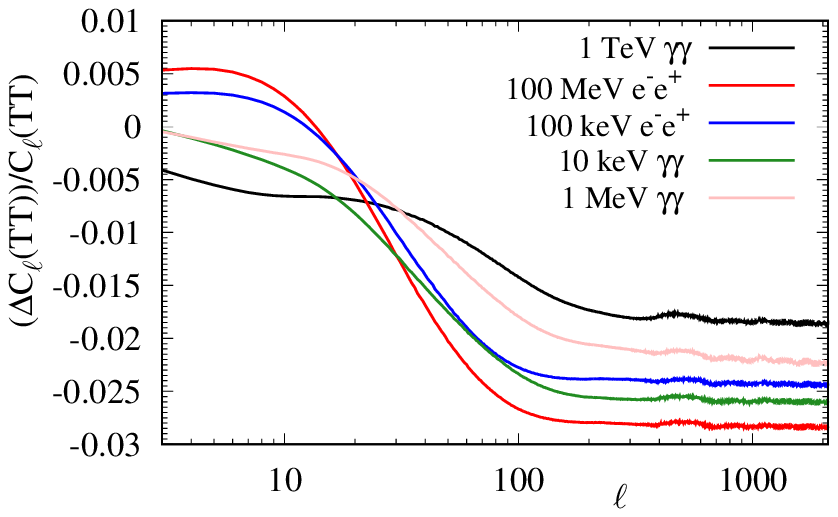}
 \caption{Fractional change in CMB temperature power spectrum for
     different dark matter decay scenarios for $\tau_X=10^{14}~{\rm s}$.}
   \label{fig:postTT}
\end{figure}

In this section, we show for completeness, the change in recombination
histories for dark matter decaying during recombination
($\tau_X=10^{13}~{\rm s}$) and before recombination ($\tau_X=10^{12}~{\rm
  s}$) in Fig. \ref{fig:xe}. We also give the corresponding change in the CMB temperature and
polarization power spectrum in Fig. \ref{fig:ps}. The fractional change in
the temperature power spectrum for post recombination decay is plotted in
Fig. \ref{fig:postTT}, see section \ref{subsec:exc} for discussion. For all
plots 2-$\sigma$ upper limits of $f_X$ as derived in Sec. \ref{sec:calc}
are used.

 The dominant effect of energy injection is to damp the CMB anisotropy
 power spectrum at high $\ell$. For lifetimes longer than recombination epoch ($\tau_X=10^{14}$s), freeze-out ionization fraction is most affected which
 results in step-function like damping. For lower lifetimes, the recombination
 history is modified closer to the hydrogen recombination epoch also and
 results in the  $\ell$-dependent damping in power spectra.

\section{Effect of dark matter decay on the 6 $\Lambda$CDM parameters}\label{app:deg}
 \begin{figure}[!tbp]
   \begin{subfigure}[b]{0.4\textwidth}
    \includegraphics[scale=0.4]{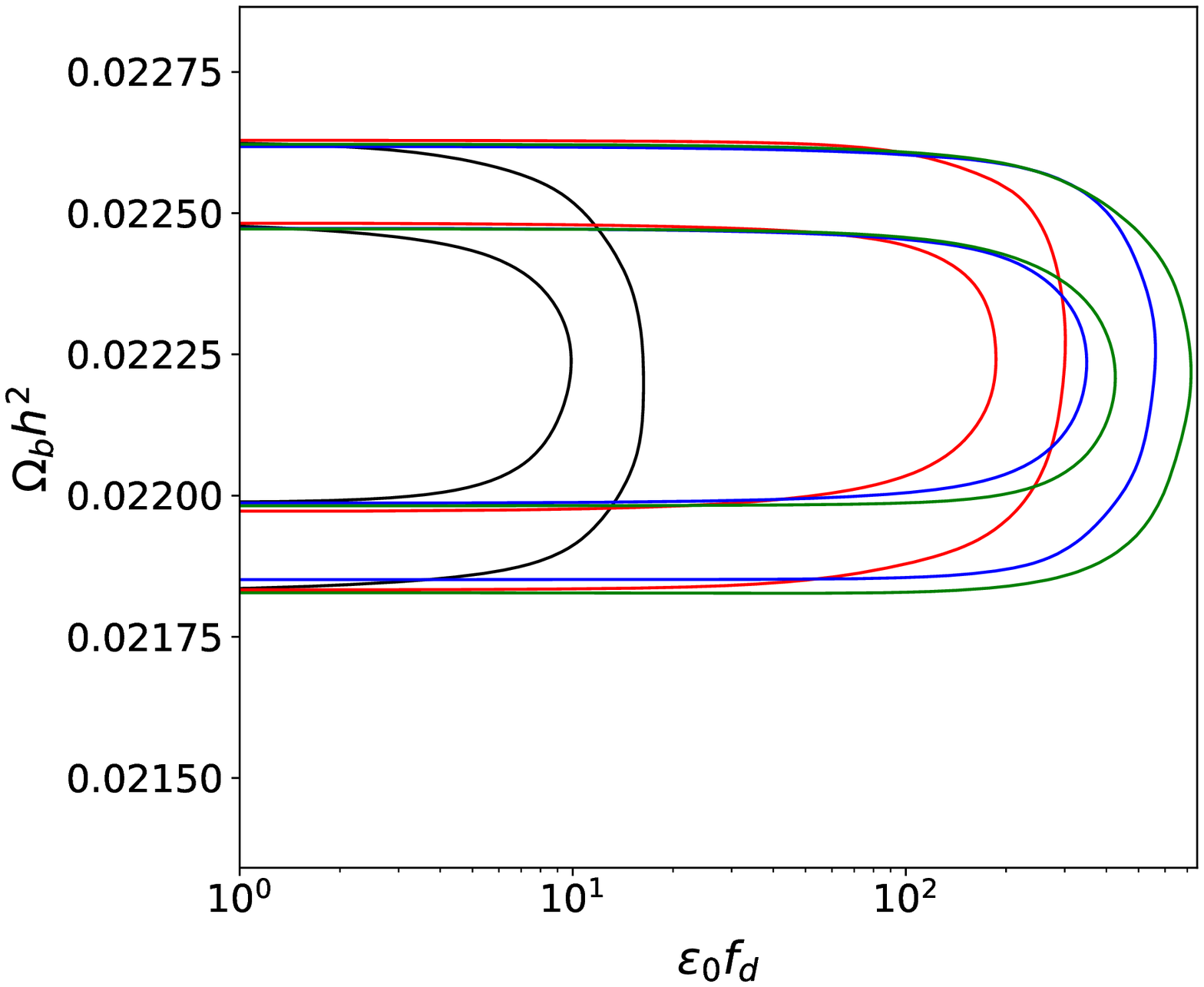}
     \label{fig:fvsb}
  \end{subfigure}\hspace{65 pt}
  \begin{subfigure}[b]{0.4\textwidth}
    \includegraphics[scale=0.4]{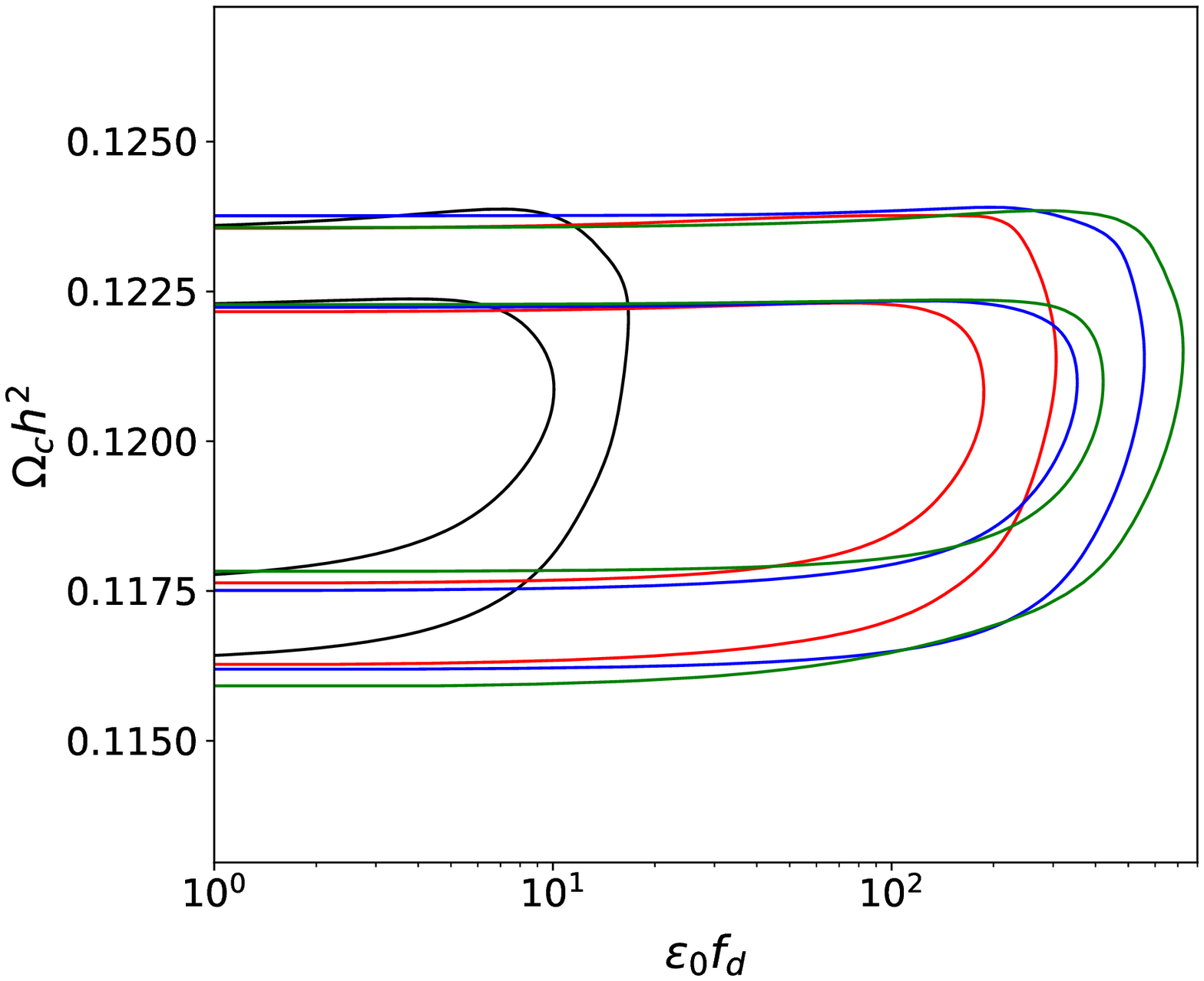}
    \label{fig:fvsc}
    \end{subfigure}\\
    
    \begin{subfigure}[b]{0.4\textwidth}
    \includegraphics[scale=0.4]{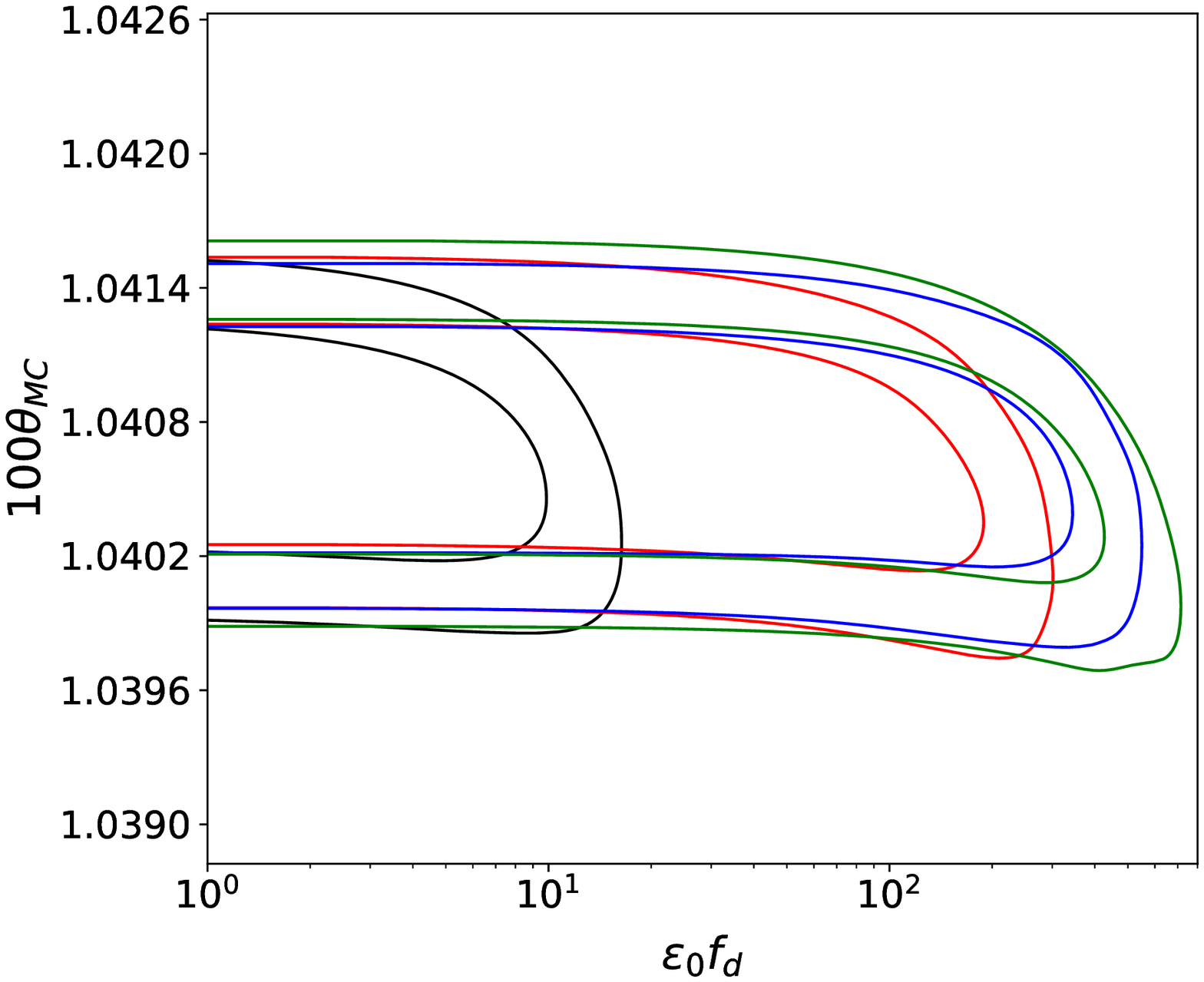}
  \label{fig:fvstheta}  
  \end{subfigure}\hspace{65 pt}
  \begin{subfigure}[b]{0.4\textwidth}
    \includegraphics[scale=0.4]{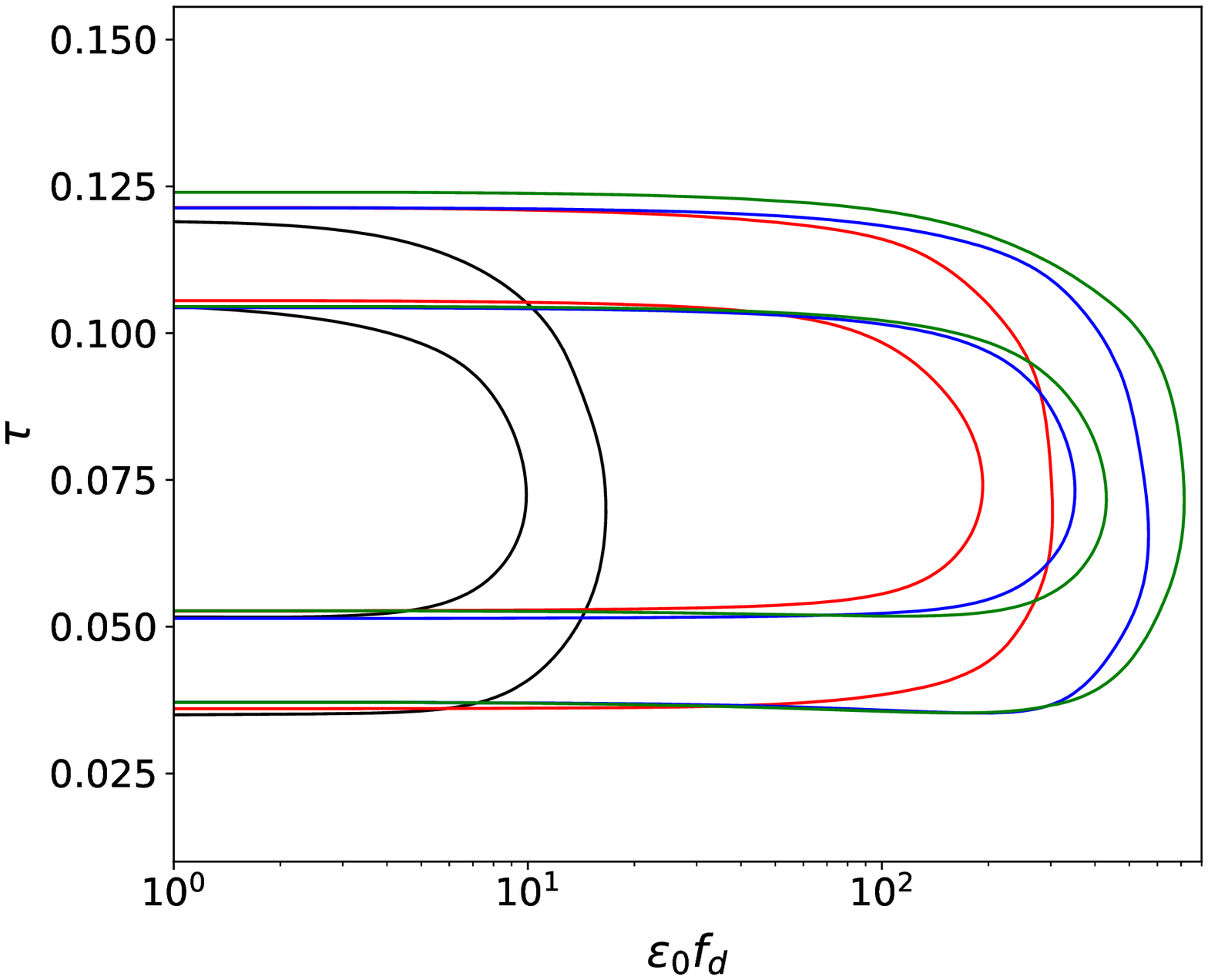}
   \label{fig:fvstau}   
   \end{subfigure}
   \begin{subfigure}[b]{0.4\textwidth}
    \includegraphics[scale=0.4]{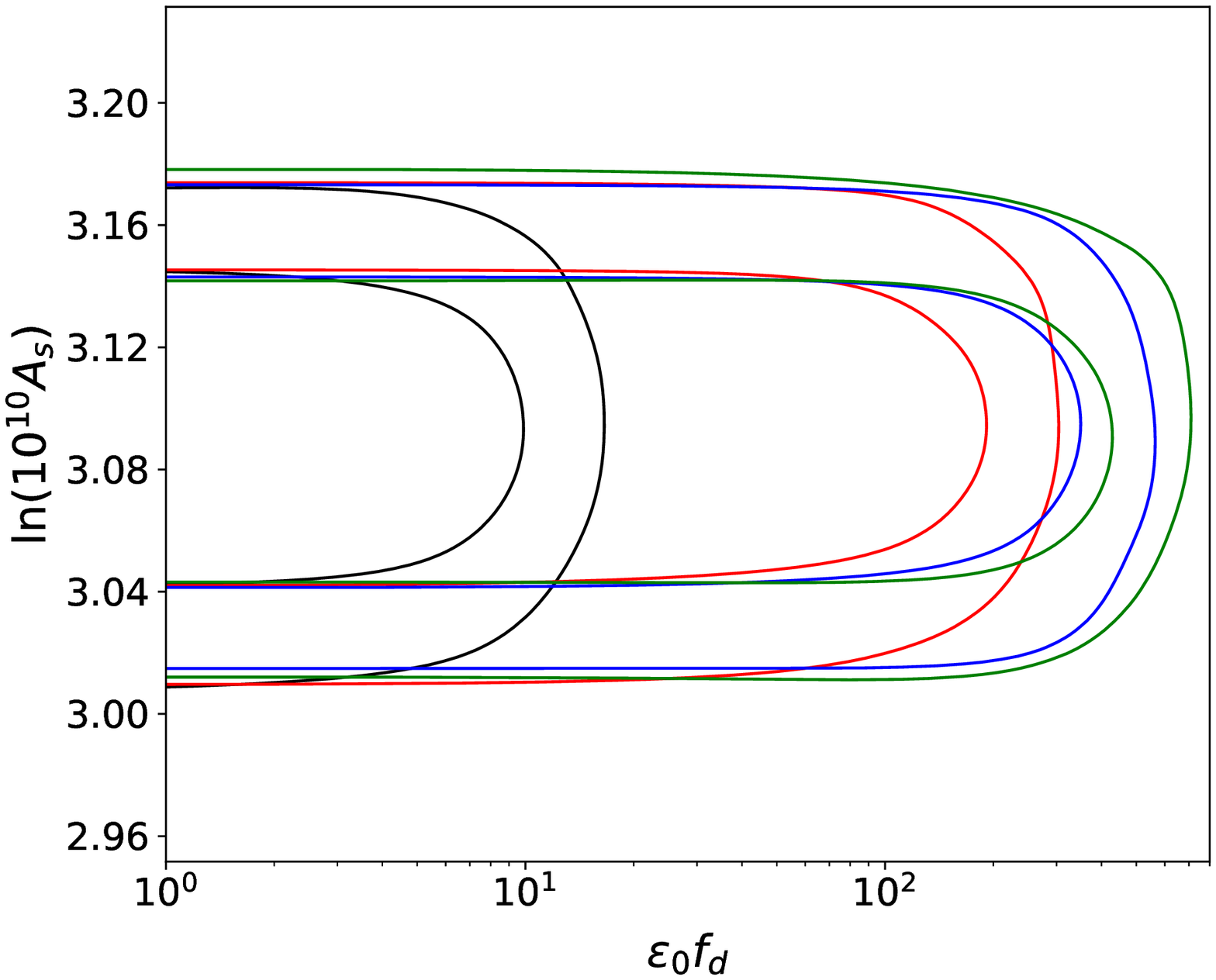}
  \label{fig:fvsAs}  
  \end{subfigure}\hspace{65 pt}
  \begin{subfigure}[b]{0.4\textwidth}
    \includegraphics[scale=0.4]{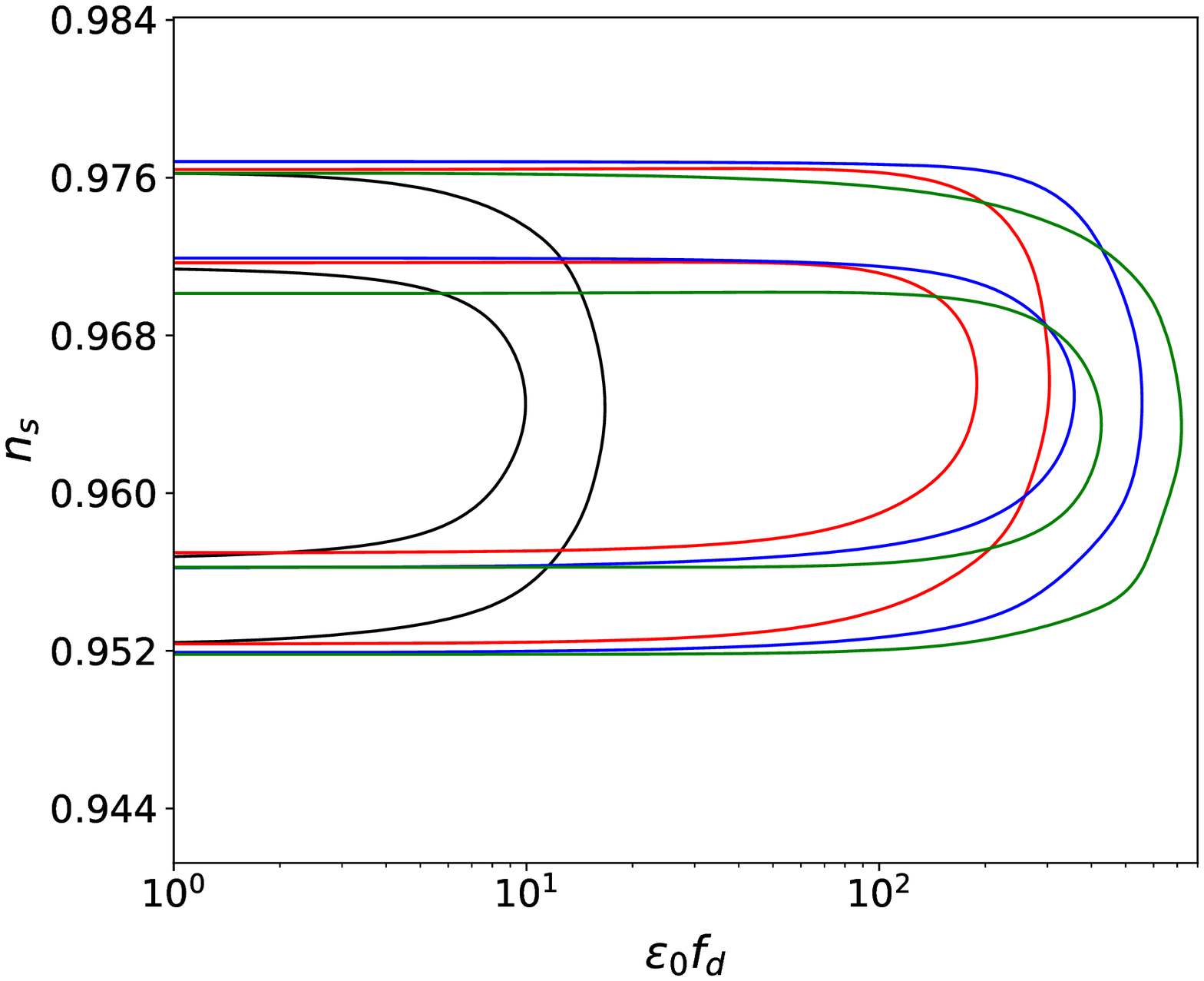}
   \label{fig:fvsns}   
   \end{subfigure}
   \caption{1-$\sigma$, 2-$\sigma$ contour for $f_X$ and 6 $\Lambda$CDM parameters for different energy injection scenarios (1 TeV $\gamma \gamma$ (black), 100 MeV $e^-e^+$ (red), 100 keV $e^-e^+$ (blue), 1 MeV $\gamma \gamma$ (green) with $\tau_X$=$10^{12}$s. $f_X$ is related to $\epsilon_0 f_d$ as $f_X$=$\epsilon_0 f_d \times 10^{-10}$. }
   \label{fig:Cl3}
   \end{figure}

 The 2-parameter 1-$\sigma$ and 2-$\sigma$ contours
are plotted in  Fig. \ref{fig:Cl3}.  We see that there is almost no
degeneracy between the $\Lambda$CDM  parameters and energy injection from
dark  matter decay. This is consistent with the previous results of
\cite{SC2017} who consider $\tau_X\ge 10^{13}~{\rm s}$. As a result there is negligible change in the 6 $\Lambda$CDM cosmological
parameters, the baryon density ($\Omega_{\rm b}h^2$), the cold dark matter
density  ($\Omega_{\rm c}h^2$), the angular acoustic scale at recombination
($\theta_{\mathrm{MC}}$), the optical depth to reionization ($\tau$), and the
amplitude ($A_{\rm s}$) and spectral index of ($n_{\rm s}$) of primordial
fluctuations. This can be seen in  Table \ref{tab:table3} where  we give the mean and 1-$\sigma$ deviation for
standard cosmological parameters for  different energy injection scenarios
and as well as Planck 2015 \cite{Pl2015} and Planck 2018 \cite{Pl2018}
$\Lambda$CDM parameter without any energy injection. The main change from
2015 to 2018 is the improved low $\ell$ polarization data which changes the
reionization optical depth. However, since there is no degeneracy between
energy injection and $\tau$, we do not expect our constraints, which were
derived with 2015 likelihoods, to be affected.
   \par
   \hspace{1cm}

\begin{table}[h!]
\fontsize{9}{9}\selectfont 
\caption{Mean values with 1-$\sigma$ deviation for standard cosmological
  parameters with varying energy and lifetime compared to the Planck 2015
  \cite{Pl2015} and Planck 2018 \cite{Pl2018}  $\Lambda$CDM results without energy injection.}
   \label{tab:table3}
  \begin{center}  
    \begin{tabular}{l|c|r|l|c|r|l} 
     Case ($\tau_X$) & $\Omega_bh^2$ & $\Omega_ch^2$ & 100$\theta_{MC}$ & $\tau$ & ln($10^{10}A_s$) & $n_s$   \\
      \hline
    PL2015 & 0.02225$\overset{+}{-}$0.00016 & 0.1198$\overset{+}{-}$0.0015 & 1.04077$\overset{+}{-}$0.00032 & 0.079$\overset{+}{-}$0.017 & 3.094$\overset{+}{-}$0.034 & 0.9645$\overset{+}{-}$0.0049 \\
    PL2018 & 0.02236$\overset{+}{-}$0.00015 & 0.1202$\overset{+}{-}$0.0014 & 1.04090$\overset{+}{-}$0.00031 & $0.0544^{+0.0070}_{-0.0081}$ & 3.045$\overset{+}{-}$0.016 & 0.9649$\overset{+}{-}$0.0044 \\
    1 TeV ($10^{16}s$) & 0.02221$\overset{+}{-}$0.00016 & 0.1204$\overset{+}{-}$0.0015 & 1.04066$\overset{+}{-}$0.00032 & 0.074$\overset{+}{-}$0.017 & 3.092$\overset{+}{-}$0.032 & 0.9643$\overset{+}{-}$0.0049 \\
   1 TeV ($10^{14}s$) & 0.02222$\overset{+}{-}$0.00016 & 0.1203$\overset{+}{-}$0.0015 & 1.04067$\overset{+}{-}$0.00033 & 0.075$\overset{+}{-}$0.017 & 3.093$\overset{+}{-}$0.033 & 0.9636$\overset{+}{-}$0.0048 \\ 
  1 TeV ($10^{13}s$) & 0.02221$\overset{+}{-}$0.00017 & 0.1205$\overset{+}{-}$0.0019 & 1.04063$\overset{+}{-}$0.00039 & 0.075$\overset{+}{-}$0.018 & 3.094$\overset{+}{-}$0.033 & 0.9638$\overset{+}{-}$0.0049 \\
      1 TeV ($10^{12}s$) & 0.02222$\overset{+}{-}$0.00016 & 0.1204$\overset{+}{-}$0.0015 & 1.04067$\overset{+}{-}$0.00034 & 0.075$\overset{+}{-}$0.017 & 3.093$\overset{+}{-}$0.033 & 0.9643$\overset{+}{-}$0.0048 \\
      100 MeV ($10^{16}s$) & 0.02222$\overset{+}{-}$0.00016 & 0.1202$\overset{+}{-}$0.0015 & 1.04070$\overset{+}{-}$0.00033 & 0.072$\overset{+}{-}$0.017 & 3.097$\overset{+}{-}$0.032 & 0.9635$\overset{+}{-}$0.0048 \\
      100 MeV ($10^{14}s$) & 0.02219$\overset{+}{-}$0.00021 & 0.1208$\overset{+}{-}$0.0026 & 1.04061$\overset{+}{-}$0.00045 & 0.070$\overset{+}{-}$0.020 & 3.094$\overset{+}{-}$0.034 & 0.9631$\overset{+}{-}$0.0051 \\
      100 MeV ($10^{13}s$) & 0.02221$\overset{+}{-}$0.00016 & 0.1205$\overset{+}{-}$0.0016 & 1.04064$\overset{+}{-}$0.00034 & 0.075$\overset{+}{-}$0.017 & 3.092$\overset{+}{-}$0.033 & 0.9639$\overset{+}{-}$0.0048 \\
      100 MeV ($10^{12}s$) & 0.02222$\overset{+}{-}$0.00016 & 0.1204$\overset{+}{-}$0.0015 & 1.04054$\overset{+}{-}$0.00035 & 0.077$\overset{+}{-}$0.017 & 3.095$\overset{+}{-}$0.033 & 0.9650$\overset{+}{-}$0.0048 \\
      1 MeV ($10^{16}s$) & 0.02222$\overset{+}{-}$0.00016 & 0.1203$\overset{+}{-}$0.0015 & 1.04065$\overset{+}{-}$0.00033 & 0.073$\overset{+}{-}$0.018 & 3.094$\overset{+}{-}$0.033 & 0.9635$\overset{+}{-}$0.0048 \\
      1 MeV ($10^{14}s$) & 0.02222$\overset{+}{-}$0.00016 & 0.1204$\overset{+}{-}$0.0016 & 1.04064$\overset{+}{-}$0.00035 & 0.074$\overset{+}{-}$0.018 & 3.095$\overset{+}{-}$0.032 & 0.9636$\overset{+}{-}$0.0049 \\
      1 MeV ($10^{13}s$) & 0.02222$\overset{+}{-}$0.00016 & 0.1205$\overset{+}{-}$0.0016 & 1.04061$\overset{+}{-}$0.00036 & 0.077$\overset{+}{-}$0.018 & 3.095$\overset{+}{-}$0.034 & 0.9643$\overset{+}{-}$0.0048 \\
      1 MeV ($10^{12}s$) & 0.02221$\overset{+}{-}$0.00016 & 0.1204$\overset{+}{-}$0.0015 & 1.04051$\overset{+}{-}$0.00037 & 0.076$\overset{+}{-}$0.017 & 3.092$\overset{+}{-}$0.033 & 0.9634$\overset{+}{-}$0.0047
    \end{tabular}
  \end{center}
   \end{table}

\end{document}